\documentclass[onecolumn]{revtex4}
\usepackage{graphicx}
\usepackage[]{epsfig}
\usepackage{amsmath}
\usepackage{color}
\usepackage{bm}
\usepackage{cancel}
\newcommand{\beq}{\begin{equation}}
\newcommand{\eeq}{\end{equation}}
\newcommand{\beqd}{\begin{displaymath}}
\newcommand{\eeqd}{\end{displaymath}}
\newcommand{\beqa}{\begin{eqnarray}}
\newcommand{\eeqa}{\end{eqnarray}}

\newcommand{\sign}{{\rm sign}}

\newcommand{\dQ}{\delta Q}

\newcommand{\comment}[1]{}

\newcommand{\Q}{\tilde{Q}}

\newcommand{\Tr}{{\rm Tr}\,}

\newcommand{\W}{{\mathcal W}}
\newcommand{\Y}{{\mathcal Y}}
\newcommand{\y}{\upsilon}

\begin{document}
\title{Logarithmic critical slowing down in complex systems: from statics to dynamics}

\author{L. Leuzzi$^{1,2}$ and T. Rizzo$^{3,2,*}$} 

\affiliation{$^1$ Institute of Nanotechnology of the National Research Council of Italy, CNR-NANOTEC, Rome Unit, Piazzale A. Moro 5, I-00185, Rome, Italy \\ 
$^2$ Physics Dept., Sapienza University, Piazzale A. Moro 5, I-00185, Rome, Italy \\ $^3$ Institute of Complex Systems of the National Research Council of Italy, CNR-ISC, Sapienza Roma Unit, Piazzale A. Moro 5, I-00185, Rome, Italy}
\email{tommaso.rizzo@cnr.it}


\begin{abstract}
We consider second-order phase transitions in which the order parameter is a replicated overlap matrix. We focus on a tricritical point that occurs in a variety of mean-field models and that, more generically, describes higher order liquid-liquid or liquid-glass transitions.  We show that the static replicated theory implies slowing down with a logarithmic decay in time. 
The dynamical equations turn out to be those predicted by schematic Mode Coupling Theory for supercooled viscous liquids at  a $A_3$ singularity, where the parameter exponent is $\lambda=1$.
We obtain a quantitative expression for the parameter $\mu$ of the logarithmic decay in terms of cumulants of the overlap, which are physically  observable in experiments or numerical simulations. 
\end{abstract}
\maketitle
\section{Introduction}

In the present work we study a peculiar kind of  critical slowing down occurring in the dynamics of slowly relaxing complex glassy systems, in which the correlation function of the relevant dynamic variables decays logarithmically in time. 
This is different from the usual behavior of, e. g., the correlation function of density fluctuations in supercooled liquid next to the dynamic arrest occurring in mean-field theories for glasses, somehow describing  the real-world (off-equilibrium) glass transition of liquid glass-formers. 
In that case, the correlator next to the transition displays a two step behavior: towards a plateau at short times and from the plateau towards zero correlation at longer times. The plateau becoming longer and longer as the external parameters bring the system nearer to the dynamic arrest line in the phase diagram. 
In G\"otze's Mode-Coupling Theory (MCT) \cite{Goetze84,Goetze85,Goetze89,Goetze09} a dynamic arrest critical point is referred to as a $A_2$ singularity, according to the classification of Arnold's catastrophes theory. The critical point corresponding to a logarithmic decay is, instead, a $A_3$ cusp singularity, a tricritical point signalling the end-point of a liquid-liquid (or glass-glass) dynamic transition.

More in detail, the behavior in time of the correlation function $C(t)$ in the time-translational invariant (TTI) regime at the ground of MCT 
is usually characterized by an initial power-law decay $t^{-a}$  towards a constant value, often related to the $\beta$ relaxation occurring in glass formers, and by a decay $-t^b$ from the plateau of $C(t)$ as the liquid system begins approaching thermodynamic equilibrium.
The exponents $a$ and $b$ are related by the well known formula for the so-called ``parameter exponent'' 
\beq
\lambda={\Gamma^2(1-a)\over\Gamma(1-2a)}={\Gamma^2(1+b)\over\Gamma(1+2b)}\ ,
\label{lambdaintro}
\eeq
holding at the dynamical $A_2$ singularity of the MCT for undercooled viscous liquids, $\Gamma(x)$ being the gamma function. Changing the values of the external parameters  a dynamic arrest line can be drawn in the phase diagram consisting of $A_2$ points. In given systems the exponent parameter $\lambda$ tends to $1$ along the dynamic arrest line, approaching a $A_3$ point. In that limit
the exponent $a$ tends to zero and  
 logarithmic corrections  become relevant to the relaxation.

 Hereafter, we present a general method for the quantitative computation of the coefficient of the logarithmic decay of the density-density correlation functions in viscous liquids. 
G\"oetze and Sj\"ogren \cite{Goetze89c} predicted a $1/\ln t^2$ decay exactly at the $A_3$ singularity and  a behavior of the kind
$-(\ln t)^\gamma$ as $a \sim 0$, $\lambda \sim 1$.
Their exemplifying case is the $F_{13}$ mode-coupling schematic theory \cite{Goetze09}.  However, in this case, the $A_3$ singularity cannot be directly accessed in experiments or in numerical simulations because it  occurs in the region of the phase diagram pertaining to the glassy phase, beyond the dynamic arrest line  where TTI breaks down and the MCT does not hold anymore.
Therefore, in the liquid phase the presence of the $A_3$ singularity is  only felt in weakly logarithmic corrections to the power-law $\beta$ decay in regions of the space parameters (temperature, packing fraction, ...) close enough to it.
 Other systems displaying this kind of singularity include disordered spin-glass models \cite{crisanti1993spherical,Crisanti06,Crisanti15,franz2013note,cammarota2012aging}, liquids in porous media both in the MCT \cite{Krakoviack05, Krakoviack07}
and in the hypernetted-chain approximations \cite{Madden88,Given92} and liquid models with pinned particles \cite{cammarota2012ideal}. 
The $-\ln t$ behavior of the correlation function appears to be the correct fitting law for about a decade or two in most of the
known experiments and numerical simulations of repulsive colloids \cite{Sjogren91,Zaccarelli02,Sciortino03,Sciortino03b,Goetze09}.
Also cases where the $A_3$ singularity is directly accessible from the liquid phase are devised in MCT, for instance in the $F_{12}$ schematic model \cite{Goetze09}. 
The quantitative estimation of the parameters of the logarithmic behavior can, so far, be performed exactly only in MCT schematic theories. Moreover, as, e.g., in the case of the $F_{13}$ model, the $A_3$ singularities lie in the region where one of the fundamental assumptions on which MCT is built, TTI, does not hold.

In many systems the replica method offers a way to characterize  dynamical arrest phenomena in a purely static framework which is often simpler than a dynamical approach. It is, therefore, natural to assume that universal static critical properties can be obtained {\em a la} Landau from simple assumptions on the (replicated) Gibbs free energy at the corresponding critical point. 
A decade ago, it was realized that replicated theories also determine important features of critical glassy dynamics \cite{Caltagirone12a,Caltagirone12b, Caltagirone12c, Ferrari12, Parisi13a}. Notably, they give the same scale-invariant equations for the critical correlators that are often obtained by studying the actual dynamical equations of these systems. An important consequence is that the  exponent parameter $\lambda$, Eq. (\ref{lambdaintro}), can be computed in a static replicated theory.

In this paper we consider a class of replicated theories for which $\lambda=1$ and show that they predict a logarithmic decay of the correlation as obtained within MCT at the $A_3$ singularity. 
Furthermore, we show that the  coefficient describing the logarithmic decay can be quantitatively expressed  in terms of static quantities that can be measured at equilibrium and we provide the formula \eqref{muin2}, which is the most notable result of the present work.

The paper is organized as follows.
In section \ref{Outline} we present the general framework and the results.
In section \ref{DerEqua} we derive the equations for critical dynamics starting from the replicated theory. 
In section \ref{ResExp} we report the general expansion of the free energy. 
In section \ref{Sec:VertexCumul} we connect the free energy with the Gibbs free energy and we derive the main results.
In section \ref{conclusions} we give our conclusions.
In appendix \ref{App:4th} we report the $23$ fourth order vertices, as well as their associated cumulants combinations.

\section{Outline of the Results}
\label{Outline}

The framework of this paper is a Landau approach to glassiness based on replicated theories.
In a Landau approach one does not start from  any specific microscopic model, and instead: (i) identifies an order parameter, (ii) makes some assumptions on the structure of the corresponding Gibbs free energy near a critical point and (iii) explores the consequences of these assumptions. The corresponding results display a great deal of universality because the assumptions of the structure of the Gibbs free energy can be valid for many different models, often with completely different microscopic structures.  On the other hand, to be concrete, one can usually exhibit solvable mean-field models whose Gibbs free energy, as given by a first principles computation, has precisely the required structure.
Solvable models are typically obtained considering either long-range models or taking the limit of infinite dimensions, later on in this section we will mention a few mean-field models to which our general findings apply.

We will follow and expand the derivation of Ref. \cite{Parisi13a},  considering theories in which the argument of the Gibbs free energy $G(Q)$ is a replicated matrix $Q_{ab}$ with $a=1,\dots,n$. The replica number $n$ in these theories is usually continued from integer to real continuous values, and we will take into account the two important cases $n \rightarrow 0$ and $n \rightarrow 1$. Such $Q_{ab}$ matrix  naturally appears in Spin-Glasses where, due to the presence of quenched disorder, one resorts for technical reasons to the replica method. 
From a physical point of view, the order parameter to identify a glassy phase, i.e., a "multi-equilibria"  phase composed by many different states,  cannot rely on an absolute reference for a state, since no {\em a priori} clear pattern is provided because of frustration. As a consequence, the order parameter is built on   the similarity between different states. More precisely, on the whole range (hierarchy) of possible similarities, summed up in a probability distribution for the values of the overlap matrix elements.
In this case $Q_{ab}$ is naturally identified with the average of the overlap between two different replicas of the system
\beq
Q_{ab}={1 \over N}\sum_{i=1}^N \overline{\langle s_i^as_i^b\rangle} \    
\eeq
where the $s_i$, $i=1,\dots, N$, are spins.
For the case  $n\to 0$ the angle brackets are thermal averages and the overline is the average over the quenched disorder. The case $n\to 1$ applies to problems where for each disorder realization there are many metastable excited states, whose number grows exponentially with the size of the system. In this latter case, then,  the angle brackets represent thermal averages {\em inside a metastable state} and the overline represents averages over different metastable states {\em and} over the quenched disorder.  

It has been argued that a replicated order parameter may be the relevant one whenever the frozen state is amorphous because to detect symmetry-breaking we have to compare it with itself. This has led to the extension of the replica method to structural glasses \cite{Mezard96,Mezard99a,Mezard99b,Mezard99c,Mezard00} and more recently to the development of the theory of supercooled liquids in the limit of infinite dimensions \cite{parisi2020theory}.
In this context $Q_{ab}$ is naturally identified with the averaged density-density fluctuations in the momentum space in a replicated system at some wave vector $\bm k$:
\beq
\label{def:Qab}
Q_{ab} \equiv {1 \over V}{ \langle \delta\rho^*_a(\bm k)\delta \rho_b(\bm k)\rangle}. 
\eeq
where $\rho_a(k)$ is the Fourier transform 
$$\rho_a(\bm k) = \sum_{i=1}^N e^{ \imath \, \bm k \cdot  \bm r_{i}^{(a)}}$$
of the density  of $N$ particles of the replica $a$ at positions $r_{i,a}$, $i=1,\ldots,N$, $$\rho(\bm r^{(a)}) = \sum_{i=1}^N\delta\left(\bm r^{(a)}-\bm r_i^{(a)}\right)$$
and $\delta \rho_a(\bm k)$ is the fluctuation of $\rho_a(\bm k)$ with respect to its average $\langle \rho_a(\bm k)\rangle$.
We note that choice of $\bm k$ in Eq. (\ref{def:Qab}) is arbitrary and one could consider, instead, the mean-square-displacement \cite{parisi2020theory}. We refer the reader to  section II.B of \cite{rizzo2016dynamical} for a thorough discussion on the choice of the order parameter. 

In mean-field models we expect that the Gibbs free energy has a regular expansion in powers of the order parameter at the critical point, therefore we will consider the following Replica-Symmetric theory written in terms of $\delta Q_{ab} \equiv Q_{ab}-Q_c$ where $Q_c$ is the value of the order parameter at the critical point and $\delta Q_{aa}=0$:  
\begin{eqnarray}
\label{eq:Gexp}
G(\delta Q)&=& \frac{m_1}{2} \sum_{ab}^{1,n}\delta Q_{ab}^2+\frac{m_2}{2}\sum_{abc}^{1,n} \delta Q_{ab}\delta Q_{ac}
\\
\nonumber
&+&\frac{m_3}{2}\sum_{abcd}^{1,n}\delta Q_{ab}\delta Q_{cd}
\label{f:Gibbs_fe}
\nonumber
\\
&-&{w_1 \over 6}\Tr \delta Q^3-{w_2 \over 6}\sum_{ab}^{1,n}\delta Q_{ab}^3
\nonumber
\\
&-&{1 \over 24}\Bigl[y_1 \Tr \delta Q^4 + y_2 \sum_{ab}^{1,n}\delta Q_{ab}^4
\nonumber
\\
\nonumber
&&{\color{black}{+y_3 \sum_{abc}^{1,n}\delta Q_{ab}^2\delta Q_{ac}^2}} + y_4 \sum_{abc}^{1,n} \delta Q_{ab}^2\delta Q_{ac}\delta Q_{cb}
\Bigr]\ .
\end{eqnarray}
The above expression can be obtained from a microscopic description in a variety of contexts \cite{Mezard87,parisi2020theory}. In the above expression we have retained only the terms relevant for the present discussion (the $y_3$ term, actually, vanishes, as will be shown in Sec. \ref{ss:n0}), while the complete expression has actually eight third-order terms and twenty-three fourth-order terms that will be displayed later, in appendix \ref{App:4th}. At the end of section \ref{ResExp} we will explain why the other terms can be neglected. We will focus on critical points characterized by the condition $m_1=0$. 
Depending on the values of the remaining parameters and on the replica number $n$ we may have different types of transition. Three such transitions, discussed in detail in \cite{Parisi13a} are: 
\begin{itemize}
    \item[i)] 
$m_2=m_3=0$ and replica number $n\to 0$, that corresponds to a standard Spin-Glass (SG) transition in zero field or to the so-called degenerate $A_2$ singularity within MCT, 
\item[ii)]  $m_2\neq 0 \neq m_3$ and replica number $n\to 0$, that corresponds to the SG transition in a field that occurs along the de Almeida-Thouless line \cite{Mezard87,crisanti1993spherical}, and 
\item[iii)] $m_2\neq 0 \neq m_3$ and replica number $n\to 1$ that corresponds to the dynamical transition in SG systems that is  the well-known $A_2$ singularity in MCT  \cite{franz2011field,rizzo2016dynamical}.
\end{itemize}

In dynamics one is typically interested in the correlation $C(t)$ between the configuration of the system at time $t=0$ and the configuration of the system at time $t$ which is the dynamical counterpart of the two-point order parameter $Q_{ab}$.
In spin systems it is naturally defined as:
\beq
C(t) \equiv {1 \over N} \sum_{i=1}^N \overline{\langle s_i(0)s_i(t)\rangle }
\eeq
while in structural glasses it is given by
\beq
C(t) \equiv  {1 \over V} { \langle \delta\rho^*(\bm k,0)\ \delta \rho(\bm k,t)\rangle} 
\eeq
In the liquid/paramagnetic phase the function $C(t)$ decays exponentially but the correlation time diverges at the critical point.
As mentioned in the introduction it has been shown \cite{Parisi13a} that the structure of the replicated Gibbs free energy at the critical point determines  also the essential features of critical dynamics.
More precisely in the case of the SG transition (i) one can show that the TTI correlation at large time-differences $t$ is described  by:
\beq
C(t)=m_1 \, f\left(\frac{t}{t^*}\right)\ \ \ t \gg 1,\ 
\eeq
for small positive $m_1$, where the time scale $t^*$ grows like
\
$$t^* \propto {1 \over m_1^{1
    \over a}},$$
the exponent $a$ is a solution of the equation
\beq
{w_2 \over w_1}={\Gamma^2(1-a)\over\Gamma(1-2a)}\ ,
\label{w2w1a}
\eeq
and the function $f(x)$ obeys the scale invariant equation:
\begin{eqnarray} 
0\,&=&\,  f(x)\,+f^2(x)\left(1-{w_2\over w_1}\right) 
\\
\nonumber&&+\int_0^x \left[f(x-y)-f(x)\right]\dot{f}(y)dy.
\end{eqnarray}
The solution of the above equation diverges as $1/x^a$ for $ x
\rightarrow 0$ and goes exponentially to zero  for $x \rightarrow \infty$. Precisely at $m_1=0$ the correlation undergoes critical slowing down and decays as a power law with exponent $a$, rather than as an exponential. 
Similar results are obtained for transitions (ii) and (iii), as, e.g., for the SK model in a field, the $p$-spin spherical and Ising models, the Random Orthogonal model, or the Potts model \cite{Caltagirone12a,Ferrari12,Caltagirone12b,Caltagirone12c}.
In particular, for the transition of type (iii) one recovers exactly the same scale invariant equations of the critical correlators in MCT ({\it i.e.} eq. 6.55a in \cite{Goetze09}) with the parameter exponent given by:
\beq
\lambda={w_2 \over w_1} ,
\label{w2w1b}
\eeq

The above results show that critical dynamics at the three transitions considered is universal because it follows solely from the structure of the replicated Gibbs free energy. Furthermore the above relationship extends the range of predictions that the replica approach can provide. 
Besides, the connection between the replicated Gibbs free energy and the parameter exponent leads to a connection with connected correlation functions of the order parameter: 
the proper vertexes $w_2$ and $w_1$ in Eq. (\ref{eq:Gexp}) are associated to vertexes of the Free energy, function of fields in the replica space coupled to the overlap fluctuations,  which are given by the connected correlation functions of $\delta Q_{ab}$. 
For instance, one obtains that
\beq
{w_2 \over w_1}={\omega_2 \over \omega_1}
\label{womega}
\eeq 
where $\omega_1$, $\omega_2$ are six-point functions given, respectively, by:
\beq
\label{def:ome1}
\omega_1={1 \over N} \sum_{ijk}\overline{\langle s_i s_j\rangle_c \langle s_j s_k\rangle_c \langle s_k s_i\rangle_c}
\eeq
\beq
\label{def:ome2}
\omega_2={1 \over 2 N} \sum_{ijk}\overline{\langle s_i s_j s_k\rangle_c^2 } \ ,
\eeq
where the suffix $c$ stands for {\em connected} correlation functions. 
As mentioned before, we recall that for transition (i) and (ii) ($n\to 0$) the angle brackets in the above expressions stand for thermal averages and the overline stands for the average over the quenched disorder. 
For transition (iii) ($n\to 1$) the angle brackets in the above expression stand for thermal averages {\it inside} a metastable state and the overline stands for the average over the different metastable states and over the quenched disorder.

In this paper we extend the above analysis to a class of critical points characterized by a replicated Gibbs free energy of the  type (\ref{eq:Gexp}) but with $w_2=w_1 \equiv w $, i. e., with $\lambda=1$. To be definite let us give a few examples of solvable mean-field models that display such a transition.
Let us consider the most general fully connected Spin-Glass models with multi-$p$-spin interactions:
\begin{equation}
    \mathcal{H} = - \sum_p \, \sum_{i_1< i_2 <\dots <i_p}  J_{i_1\dots i_p}s_{i_1}\dots s_{i_p}
\end{equation}
where the $J's$ are quenched random interactions and the $s_i$ can be either Ising spin or satisfy a spherical constraint.
In the spherical $3$-spin case in presence of a magnetic field there is a tricritical point with $\lambda=1$ in the temperature-magnetic field plane where a line of discontinuous transitions meets a line of continuous transitions \cite{crisanti1992spherical,crisanti1993spherical}.   
Another example is that of a mixed $2+3$ model that corresponds to the so-called schematic $F_{12}$ model in the context of MCT. 
In the phase diagram, e.g., in the plane of the magnitudes of the $2$-spin and the $3$-spin interaction  there is a $\lambda=1$ critical point. Upon increasing the relative magnitude of the $3$-spin interaction, a line of continuous transitions meets a line of discontinuous transitions \cite{franz2013note,caltagirone2013critical}. Random pinning of a spherical $p$-spin glass  model, {\it  i.e.} freezing a  fraction $c$ of the spins \cite{cammarota2012ideal,cammarota2012aging}, is also relevant: in the temperature-concentration plane there is a line of discontinuous transitions that, upon increasing the concentration, ends in a point characterized by $\lambda=1$.
Finally we mention the Potts Spin-Glass with Hamiltonian:
\begin{equation}
 \mathcal{H} = - \sum_{i<j} \, J_{ij} \, (p \, \delta_{s_i s_j}-1)   
\end{equation}
where the $s_i$ are Potts spins with $p$ states and $J_{ij}$ is a quenched random interaction. For $p \leq 4$ it displays a continuous SG transition characterized by $\lambda = (p-2)/2$, which implies $\lambda=1$ for $p=4$, in both the fully connected \cite{gross1985mean,Caltagirone12c} and the finite-connectivity case \cite{goldschmidt1988potts}.

If $\lambda=1$ the correlation cannot decay with a power-law because eqs. (\ref{w2w1b}) or (\ref{lambdaintro}) yield $a=b=0$. 
Indeed for all the three types of transitions we will show that, at large times:
\beq
C(t) - C(\infty) = \ {2 \pi^2 \over 3\, \mu \ln^2 (t/t_1)} +{24 \zeta(3) \over \mu \ln^3 (t/t_1)}\ln \ln (t/t_1)+\dots
\label{clog}
\eeq
where $\zeta(i)$ is the Riemann's $\zeta$-function and $t_1$ is an unkwown time scale that cannot be determined due to the time-scale invariance of equation.
We note that this expression was obtained by G\"otze and Sj\"ogren \cite{Goetze89c} within MCT in the context of the so-called $A_3$ singularities \cite{Goetze09} and, indeed, we will derive the same dynamical equations.
The parameter $\mu$ in Eq. (\ref{clog}) depends on the quartic coupling constants of the Replicated Gibbs Free energy (\ref{eq:Gexp}) through:
\beq
\mu=-{ y_1+y_2-y_4\over 3 w} \ .
\label{muin1}
\eeq
As usual, the coefficients of the Gibbs free energy can be expressed in terms of 
  four-point connected correlation functions of the order parameter and, thus, we will show that the parameter $\mu$ can be calculated in terms of physical measurable observables as
\begin{equation}
\mu =  - \frac{r}{3 \omega}\left(\y_1+\y_2-\y_4\right)
\label{muin2}
\end{equation}
where $r \equiv \chi_{SG}^{-1}$, $\chi_{SG}$ being the so-called Spin-Glass susceptibility:
\beq
\chi_{SG} \equiv {1\over N} \sum_{ij} \overline{\langle s_is_j\rangle^2_c} \ ,
\eeq
whereas $\omega$ is either given by $\omega_1$ or $\omega_2$ defined in Eqs. (\ref{def:ome1},\ref{def:ome2}) since they are equal at the critical point that we are considering.
Eventually, the  $\y$'s are the fourth-order analogs of the $\omega$'s. As we will show in Sec. \ref{ResExp}, their expressions turn out to be:
\beqa
\y_1 & \equiv & {3 \over N} \sum_{ijkl}\overline{\langle s_i s_j\rangle_c \langle s_j s_k\rangle_c \langle s_k s_l\rangle_c\langle s_l s_i\rangle_c}
\label{intro:y1}
\\
\y_2 & \equiv  & {1 \over 2 N} \sum_{ijkl}\overline{\langle s_i s_j s_k s_l\rangle_c^2 }
\label{intro:y2}
\\
\y_4 & \equiv & {6 \over  N}  \sum_{ijk}\overline{\langle s_i s_j s_k\rangle_c \langle s_i s_j s_l \rangle_c \langle s_l s_k \rangle_c }
\label{intro:y4}
\eeqa
We will also consider the critical behavior of the physical susceptibilities. In particular, we will show that close to the critical point, where $r$ vanishes linearly with the external parameters (in mean-field models), the three-point susceptibilities $\omega_i$, $i=1,2$, diverge as 
\beq
\omega_i = \frac{w_i}{r^3} 
\eeq
and the four-point susceptibilities $\y_i$, $i=1,\dots,4$,  diverge as 
\beq
\y_i = O\left(\frac{1}{r^{5}}\right) \ .
\eeq
However, the linear combination $\y_1+\y_2-\y_4$ associated to $\mu$ is less divergent if $w_1=w_2$, as it turns out to obey the following relationship:
\beq
\y_1+\y_2-\y_4 = 6\ \frac{(w_1-w_2)^2}{r^5} + \frac{y_1+y_2-y_4}{r^4}
\label{mucrit}
\eeq
Equations (\ref{clog}), (\ref{muin1}), (\ref{muin2}) and (\ref{mucrit}) are the main results of this paper and will be derived in the following. In particular, Eqs. (\ref{clog}) and (\ref{muin1}) will be derived in the following section. In section \ref{ResExp}
the free energy will be introduced, and  the expression of its coefficients (\ref{intro:y1}-\ref{intro:y4}) will be derived. Eventually, Eq. (\ref{muin2}) will be derived in section  \ref{Sec:VertexCumul}.

\section{Derivation of the Equations of Critical Dynamics}
\label{DerEqua}

In this section we show how the expression (\ref{clog}) can be derived from the static replicated Gibbs free energy (\ref{eq:Gexp}). We first differentiate it with respect to the order parameter $\delta Q_{ab}$ obtaining the following equation of state:
\beqa
0 & = & w_1 (\delta Q^2)_{ab}+w_2 \delta Q^2_{ab}
\label{eq:Gibbs_verteces}\\
\nonumber
& + & {y_1 \over 3} (\delta Q^3)_{ab}+{y_2 \over 3}\delta Q^3_{ab}
\\
\nonumber
&+&{\color{black}{{y_3 \over 6} \delta Q_{ab}\left[(\delta Q^2)_{aa}+(\delta Q^2)_{bb}\right]+}}{y_4 \over 6}\delta Q_{ab} (\delta Q^2)_{ab}
\\
\nonumber
&+&{y_4 \over 12} \sum_c\left[\delta Q_{ac}^2 \delta Q_{cb}
+\delta Q_{bc}^2 \delta Q_{ca}\right]
\eeqa
Now  we translate the above equation into an equation for the dynamical correlation valid at large times and near the critical point. We will briefly sketch the arguments leading to this mapping but we refer the reader to Ref. \cite{Parisi13a}  for all the details of the procedure. 
The result is obtained in the context of a super-field formulation of dynamics \cite{kurchan1992supersymmetry} in which both the dynamical correlation and response functions
are represented by a single dynamical order parameter $Q(1,2)$ in terms of (commuting) times $t_{1,2}$ and Grassmannian anticommuting variables $\theta_{1,2}$, $\bar\theta_{1,2}$.

\subsection{Theories with $n=0$}
\label{ss:n0}
At equilibrium $Q(1,2)$ can be parameterized by a single time translational invariant correlation function $C(t)=C(-t)$ according to the following form that encodes causality and the Fluctuation-Dissipation-Theorem (FDT):
\beq
Q(1,2)=\left\{   1+ 
\Theta_{12}
{\partial \over \partial t_1}\right\}C(t_1-t_2)
\label{SUS}
\eeq  
with
\beqa
\Theta_{12}&\equiv &
{1 \over 2}
(\overline{\theta}_1-\overline{\theta}_2)
\left[\theta_1+\theta_2-(\theta_1-\theta_2)\ \sign(t_1-t_2)
\right]
\nonumber
\\
&=&\left\{\begin{array}{c c} (\bar\theta_1-\bar\theta_2)\theta_2 &, \quad t_1>t_2
\\
& \vspace*{-.2cm}
\\
(\bar\theta_1-\bar\theta_2)\theta_1 &,\quad t_2>t_1 .
\end{array}\right. 
\eeqa
We note that this representation is appropriate for the phase transitions characterized by a replicated free energy with $n=0$, whereas for $n=1$ a different representation must be considered (\cite{Parisi13a}, Sec. III.D). We postpone the discussion of this case to the end of this section. 
On general grounds it is to be expected that the dynamical order parameter $Q(1,2)$ at large times must be related to the static order parameter $Q_{ab}$.
Indeed, as noted in \cite{kurchan1992supersymmetry}, the static result is obtained in the so-called Fast Motion (FM) limit that corresponds to an infinitely fast microscopic dynamics. In this limit, configurations at different times are completely uncorrelated and they are equivalent to different replicas of the same system. As a consequence, in this limit $Q_{FM}(1,2)$ has a diagonal structure: 
\beq
Q_{FM}(1,2)=\delta(1,2) \left[C(0)-C(\infty)\right]+C(\infty)\ ,
\eeq 
where $\delta(1,2)$ is a delta function in the super-variables and $C(0)$, $C(\infty)$ are the values of the correlation at zero and infinite time,  respectively. 
It is useful to describe the dynamics at large but not infinite times in terms of the deviation of $Q(1,2)$ from its FM limit introducing the quantity 
\beq
\delta Q(1,2)=Q(1,2)-Q_{FM}(1,2) \ .
\eeq
The dynamical equations for $Q(1,2)$ can be obtained from a dynamical Gibbs free energy and one may  expect that the critical dynamics is determined {\it a la} Landau by its expansion in powers of $\delta Q(1,2)$. 
In \cite{Parisi13a} it is argued that the dynamical Gibbs free energy {\it must have the same structure of the replicated Gibbs free energy with the  same coupling constants} 
and, therefore, Eq. (\ref{eq:Gibbs_verteces}) translates into an identical equation for  $\delta Q(1,2)$. 

In the following we will rewrite  Eq. (\ref{eq:Gibbs_verteces}) with $\delta Q_{ab} \rightarrow \delta Q(1,2)$ as an equation for $C(t)$. In order to simplify the computation we observe that all the terms are obtained from $\delta Q(1,2)$ through the operation of exponentiation of matrix elements and dot products.
These operations preserve supersymmetry, time reversal, zero ghost number and causality (see \cite{kurchan1992supersymmetry}, section 5.5) and, therefore, their result can still be written in the form (\ref{SUS}) which is the most general form satisfying these properties. Using an appropriate even function $A(\tau)$, 
the generic exponentiation corresponds to a simple power: 
\beq
A(1,2)^k = \left\{   1+ \Theta_{12}{\partial
\over \partial t_1}\right\}A^k(t_1-t_2)
\label{Aprod}
\eeq
The dot product corresponds  to:
\beq
\int A(1,3)B(3,2)\ d3 =\left\{   1+ \Theta_{12}{\partial
\over \partial t_1}\right\}[AB](t_1-t_2) 
\label{AB12}
\eeq
where the function $[AB](t)$ stands for:
\beqa
[AB](t)&=&A(t)B(0)+B(t) A(0)
\label{ABt}
\\
\nonumber
&& -A(\infty)B(-\infty)-{d \over dt} \int_0^{t} A(t-y)B(y)dy .
\eeqa
One can check that if both $A(t)$ and $B(t)$ are even functions, then $[AB](t)$ is even and
$[AB](t)=[BA](t)$.
We recall that $\delta Q(1,2)$ is also of the form (\ref{SUS}) with $\delta C(t)=C(t)-C_{FM}(t)$ where $C_{FM}(t)$ obeys $C_{FM}(0)=C(0)$ and $C_{FM}(0^+)=C(\infty)$.  Therefore, by construction we have that $\delta C(0)=\delta C(\infty)=0$ and this simplifies considerably the evaluation of the various terms.
Using the two rules $(\ref{Aprod})$ and $(\ref{AB12})$ we can translate all the terms of Eq. \ref{eq:Gibbs_verteces} into  expressions of their dynamical counterparts.
For the quadratic terms we have:
\beqa
w_1 (\delta Q^2)_{ab} & \rightarrow &  -w_1 \ {d \over dt} \int_0^{t} \delta C(t-y)\ \delta C(y)\ dy
\\
w_2 \ \delta Q^2_{ab} & \rightarrow  & w_2 \ \delta C^2(t) .
\eeqa
In order to make contact with the MCT notation of reference \cite{Goetze89c} we define:
\beq
\delta C(t) \equiv G(t)
\eeq
and introduce the  Laplace transform of the time functions as \cite{Goetze09}:
\beq
\hat A(z) = LT[A(t)](z)\equiv i \int_0^\infty A(t)\ e^{i z t} dt \qquad , \quad \mbox{Im}[z]>0
.\eeq
The formulas for the transforms of the convolution  and of the time derivative are repeatedly used in the following derivation and we write them explicitly 
\beq
LT\left[ \int_0^t A(t-y)B(y)dy\right]=-i \hat A(z) \hat B(z) 
\label{conv}
\eeq
\beq
 LT\left[ {d A(t) \over dt}\right]=-i z \hat A(z) .
\label{deri}
\eeq
We also define $(G^2)(t)$ as the time derivative of the convolution, cf. Eq. (\ref{ABt}):
\beq
(G^2)(t)\equiv -{d \over dt} \int_0^{t} G(t-y)\ G(y)\ dy \ .
\eeq
The contributions of the quadratic terms can now be expressed as:
\beqa
w_1 \ (\delta Q^2)_{ab} & \rightarrow &  w_1 \ z \ \hat G^2(z)\ ,
\label{m1}
\\
w_2\  \delta Q^2_{ab} & \rightarrow  & w_2 \ LT[G^2(t)] \ ,
\label{m2}
\eeqa
while those of the cubic terms can be expressed as:
\beqa
{y_1 \over 3}\  (\delta Q^3)_{ab} & \rightarrow &  {y_1 \over 3}\  z^2 \hat G^3(z)
\label{m3}
\\
{y_2 \over 3}\ \delta Q^3_{ab} & \rightarrow  & {y_2 \over 3} \ LT[G^3(t)]
\label{m4}
\\
{y_3 \over 6} \ \delta Q_{ab}((\delta Q^2)_{aa}+(\delta Q^2)_{bb}) & \rightarrow  & 0
\label{m5}
\\
{y_4 \over 6}\ \delta Q_{ab} (\delta Q^2)_{ab} & \rightarrow & {y_4 \over 6}\ LT[G(t) (G^2)(t)] 
\label{m6}
\\
{y_4 \over 12} \ \sum_c[\delta Q_{ac}^2 \delta Q_{cb}+\delta Q_{bc}^2 \delta Q_{ca}] & \rightarrow  & {y_4 \over 6} \ z\  \hat G(z)\  LT[G^2(t)] .
\label{m7}
\eeqa
We stress that, in order to compute the vanishing term proportional to $y_3$, we have used the fact that, according to Eqs. (\ref{AB12}) and (\ref{ABt}):
\beq
AB (1,1)=[AB](0)=A(0)B(0)-A(\infty)B(-\infty)\ ,
\eeq
that implies:
\beq
(\delta Q^2)_{aa} \rightarrow \delta C(0)^2 -\delta C^2(\infty)=0 \ .
\eeq
Dividing Eq. (\ref{eq:Gibbs_verteces}) by $w_1$ and multiplying by $z$, we obtain:
\beqa
0 & = & z\left\{ {w_2 \over w_1} LT[G^2(t)]+z\  \hat G^2(z)\right\}
\nonumber
\\
& + & { y_1+y_2-y_4\over 3 w_1}\ z\  LT[G^3(t)]
\nonumber
\\
& - & {2 y_1 -y_4\over 6 w_1}\  z\left\{ LT[G^3(t)]- z^2 \hat G^3(z)\right\}
\nonumber
\\
&+& {y_4\over 6 w_1} \ z\ \left\{ LT[G^3(t)]+ z \ \hat G(z) \ LT[G^2(t)]\right\}
\nonumber
\\
& + & {y_4\over 6 w_1}\  z\ \left\{ LT[G(t)(G^2)(t)]+ z^2 \ \hat G^3(z)\right\}, 
\label{eqfin}
\eeqa
where we have just rearranged the various term for later convenience. Shortening $\mu \equiv (y_1+y_2-y_4)/ (3 w_1)$,
the first two lines correspond to the equation 
\beqa
0 & = & z\left\{ {w_2 \over w_1} LT[G^2(t)]+z \ \hat G^2(z)\right\}- \mu \ z\  LT[G^3(t)]
\label{EQ-CRIT4}
\eeqa
considered by G\"oetze and Sj\"ogren \cite{Goetze89c} who showed that its solution at leading order follows the logarithmic decay of Eq. (\ref{clog}).
Our equation (\ref{eqfin}) has the same form except for the additional terms of the last three lines.
In order to characterize its solution one introduces an auxiliary function $g$ by setting $G(t)=g(\ln(t/t_1))$. 
Following \cite{Goetze89c} we introduce the variable $$y\equiv \ln\frac{1}{-i z t_1}$$ and, changing the integration variable in the Laplace transform from $t$ to $u\equiv -i z t$, we obtain the relationship
\beq
-z\, \ \  {\hat G}(z) = \int_0^\infty \, du \, e^{-u}g(y+\ln \, u) .
\eeq

The Taylor expansion around $y$ gives:
\beq
-z\, \hat G(z) = g(y) + \Gamma_1 \ g'(y)+{1 \over 2}\Gamma_2\, \ g''(y) + \dots 
\label{Taylorexp}
\eeq
where 
\beq
\Gamma_n \equiv \int_0^\infty du \, e^{-u}(\ln u)^n \ .
\eeq
It follows that for a  generic product the Laplace transforms reads   
\beqa
-z\, LT[G_1^p(t) G_2^q(t)](z) & = & g_1^p g_2^q + \Gamma_1 (p g_1' g_1^{p-1} g_2^ q + q g_2' g_2^{q-1} g_1^p)+
\nonumber
\\
& + & {1 \over 2}\Gamma_2\, ( \,  p (p-1)g_1''\, g_1^{p-2} g_2^q \ + p q g_1' g_2'\, g_1^{p-1} g_2^{q-1}\ +q (q-1)g_2''\, g_2^{q-2} g_1^p) + \dots 
\eeqa
The integrals $\Gamma_n$ can be expressed as polynomials of degree $n$ of the Euler's constant $\gamma=-\Gamma_1$,  with coefficients given by combinations of the Riemann's zeta function up to $\zeta(n)$.
Furthermore, given any two functions $G_1(t)$ and $G_2(t)$ and the corresponding functions $g_1$ and $g_2$ we have \footnote{Note that there is a typographical error in Eq. (A9) in \cite{Goetze89c}, the correct expression being the one displayed here, in Eq. (\ref{eq:G1G2})}:
\beq
\label{eq:G1G2}
-z\left\{LT\left[G_1(t)G_2(t)\right]+ z \ \hat G_1(z)\  \hat G_2(z)\right\}=\zeta(2) g_1' g_2'+\dots \ ,
\eeq
where $\zeta(2) = {\pi^2 / 6}$.
With the help of the above formulas we obtain:
\beqa
 z\big\{ LT[G^2(t)]+z \ \hat G^2(z)\big\} & = & - \zeta (2) (g')^2 + 2 (\gamma \zeta(2)+\zeta(3))g' g'' +\dots
\nonumber
\\
z LT[G^3(t)] & = & -g^3  + 3 \gamma g^2 g'+ \dots 
\nonumber
\\
 z\{ LT[G^3(t)]- z^2 \ \hat G^3(z)\} & = & O\left(g \ (g')^2\right) 
\nonumber
\\
 z\{ LT[G^3(t)]+ z \ \hat G(z)\  LT[G^2(t)]\} & = & O\left(g \ (g')^2\right)
\nonumber
\\
z\{ LT[G(t)(G^2)(t)]+ z^2 \ \hat G^3(z) \} & = & O\left(g \ (g')^2\right) \ .
\label{eqfin2}
\eeqa
To derive the above formulas we used the fact that the last three lines are all of the form $-z\{LT[G_1(t)G_2(t)]+ z \ \hat G_1(z)\ \hat  G_2(z)\}$.
At leading order, at the tricritical point $w_1=w_2$, the equation (\ref{EQ-CRIT4}) becomes
\beq
-{2 \, \pi^2 \over 3}(g')^2+\mu g^3 = 0 \ ,
\label{g_y_leading}
\eeq
whose solution is the leading order of Eq. (\ref{clog}). Since solving Eq. (\ref{g_y_leading}) one observes that at leading order $g = O(y^{-2})$, we also see that the leading corrections to the first two lines is $O(y^{-7})$, while the last three lines are $O(y^{-8})$. 
Higher order terms, $O(\delta Q^4)$, in the equation of state  would also give a contribution $O(y^{-8})$ and,  therefore, the sub-leading correction is solely determined by the sub-leading corrections to the first two lines in Eq. (\ref{eqfin}). 
The coefficient of the second line is, eventually, simply the combination $\mu$, cf. Eq. (\ref{muin1}).
The solution at leading and subleading order is, eventually, the anticipated result Eq. (\ref{clog})
\beq
G(t) = \ {2 \pi^2 \over 3\, \mu \ln^2 (t/t_1)} +{24 \zeta(3) \over \mu \ln^3 (t/t_1)}\ln \ln (t/t_1)+\dots
\nonumber
\eeq

\subsection{Theories with $n=1$}
\label{ss:n1}
We now turn to the $n=1$ case. This is relevant for the important case of the dynamic transition  in SG models with  one step of Replica-Symmetry-Breaking and in glass-forming supercooled liquids \cite{Crisanti08,franz2011field,Parisi13a,rizzo2016dynamical,parisi2020theory}.
As we mentioned before, here the dynamics has to be described in a formalism that takes into account the initial condition. The formulation is given in section III.D of \cite{Parisi13a} and a complete treatment is reported in \cite{rizzo2016dynamical}. 
We will not enter into the details hereafter and we will limit ourselves to mention that the dynamical objects involved are, once again, the super-symmetric matrices $Q(1,2)$. 
Inasmuch as we did above, these can be written in terms of a single TTI correlation function $C(t)=C(-t)$:
\beq
Q(1,2) \rightarrow C(t).
\eeq
Once again Eq. (\ref{eq:Gibbs_verteces}) can be, thus, translated into an equation for $C(t)$.
As we saw before we only need to specify the analogue for $n=1$ of Eqs. (\ref{Aprod}, \ref{AB12}), i.e., the behavior of element-wise products  $A^k(1,2)$ and dot products $\int A(1,3)B(3,2)\, d3$.
For the products we just have \cite{Parisi13a}
\beq
\label{dyn_pow_n1}
A(1,2)^k \rightarrow A^k(t_1-t_2), 
\eeq
while for the dot product we have:
\beq
\label{dyn_dot_n1}
\int A(1,3)B(3,2)\, d3 \rightarrow [AB](t_1-t_2) 
\eeq
with \cite{rizzo2016dynamical}:
\beqa
[AB](t)&=&A(t)B(0)+B(t) A(0)
\\
\nonumber
&& -{d \over dt} \int_0^{t} A(t-y)\ B(y) \ dy .
\eeqa
Note that this is {\it different} from Eq. (\ref{ABt}) for the $n=0$ dynamics, due to the absence of the term $A(\infty)B(\infty)$.
However, since we are computing products of $\delta Q(1,2)$ and we have $\delta C(0)=\delta C(\infty)=0$, the two expressions yield the same results. As a consequence  the mappings (\ref{m1}-\ref{m7})
hold for $n=1$.
We stress that again Eq. (\ref{m5}) gives a vanishing contribution as
\beq
AB (1,1)=[AB](0)=A(0)B(0) \,
\eeq
(again with no term $-A(\infty) B(\infty)$) and therefore
\beq
(\delta Q^2)_{aa} \rightarrow \delta C(0)^2 =0 \ .
\eeq
The validity of the mappings implies that Eq. (\ref{eqfin}) holds also for $n=1$ and, consequently, the same logarithmic relaxation behavior, cf. Eq. (\ref{clog}), is derived, with the same procedure  performed for $n=0$, in \ref{ss:n0}.

\section{Expansion of the Replicated Free Energy at Fourth Order}
\label{ResExp}

In this section we express the coefficients of the {\it free energy}  of a generic model in terms of physical observables, namely the cumulants of the overlap. 
For the sake of clarity  in the following we will reproduce the derivation of the third order expansion, initially reported in Ref.  \cite{Parisi13a},  while we will postpone to appendix \ref{App:4th} the derivation of the fourth order result.
For the sake of readability we might repeat some of the definitions already given before.

Averages in the replicated system can be rewritten as
\beq
\langle \cdots \rangle \equiv \overline{ \langle \cdots \rangle_J}\ ,
\eeq
where $\langle \cdots \rangle_J$ are thermal averages at fixed quenched disordered interactions $J$ while the overline is the average over the couplings that must be performed reweighting  each disorder realization with the single system partition function to the power $n$: 
\beq
\overline{O_J}=\frac{\int dP(J) O_J Z_J^n}{\int dP(J) Z_J^n}
\eeq
Note that the thermal averages between different replicas factorize prior to the disorder averages.
We define the following free energy functional:
\beq
F(\lambda) \equiv -{1 \over N}\ln \langle e^{\sum_{(ab)}N\lambda_{ab}\delta \Q_{ab}}\rangle
\label{FDEF}
\eeq
where 
\beq
\delta \Q_{ab}={1 \over N}\sum_i s_i^a s_i^b- q
\eeq
and 
\beq
q \equiv {1 \over N}\sum_{i}\langle s_i^a s_i^b \rangle = \sum_i \overline{\langle s_i\rangle_J^2}
\eeq
We note that the above free energy functional arises if we apply to each spin $s_i^a$ of each replica a Gaussian distributed random field $h_i^a$ with covariance matrix given by $\overline{h_i^a h_j^b}=\lambda_{ab}\delta_{ij}$. 
Following \cite{Parisi13a} we start by expanding $F(\lambda)$ in powers of $\lambda$ at fourth order assuming $\lambda_{aa}=0$ $\forall a$:
\beq
F(\lambda)= -{1 \over 2}\sum_{(ab),(cd)}\lambda_{ab}G_{ab,cd}\lambda_{cd}-{1\over 6} \sum_{(ab),(cd),(ef)}\W_{ab,cd,ef}\lambda_{ab}\lambda_{cd}\lambda_{ef}-{1\over 24} \sum_{(ab),(cd),(ef),(gh)}\Y_{ab,cd,ef,gh}\lambda_{ab}\lambda_{cd}\lambda_{ef}\lambda_{gh}, 
\label{eq:FEN4th}
\eeq
The $G's$, the $\mathcal W$'s and the $\mathcal Y$'s  are, respectively, the connected correlation functions of order two, three and four. 
In the Replica Symmetric case  the total number of different cumulants of order $K$,  $\mathcal C_{a_1b_1,\dots,a_K b_K}$, is given by the set of possible diagrams (connected and disconnected) with $K$ legs with the condition that any leg connects different vertices (due to the assumption $\lambda_{aa}=0$). 
\begin{figure}[t!]
\begin{center}
\epsfig{file=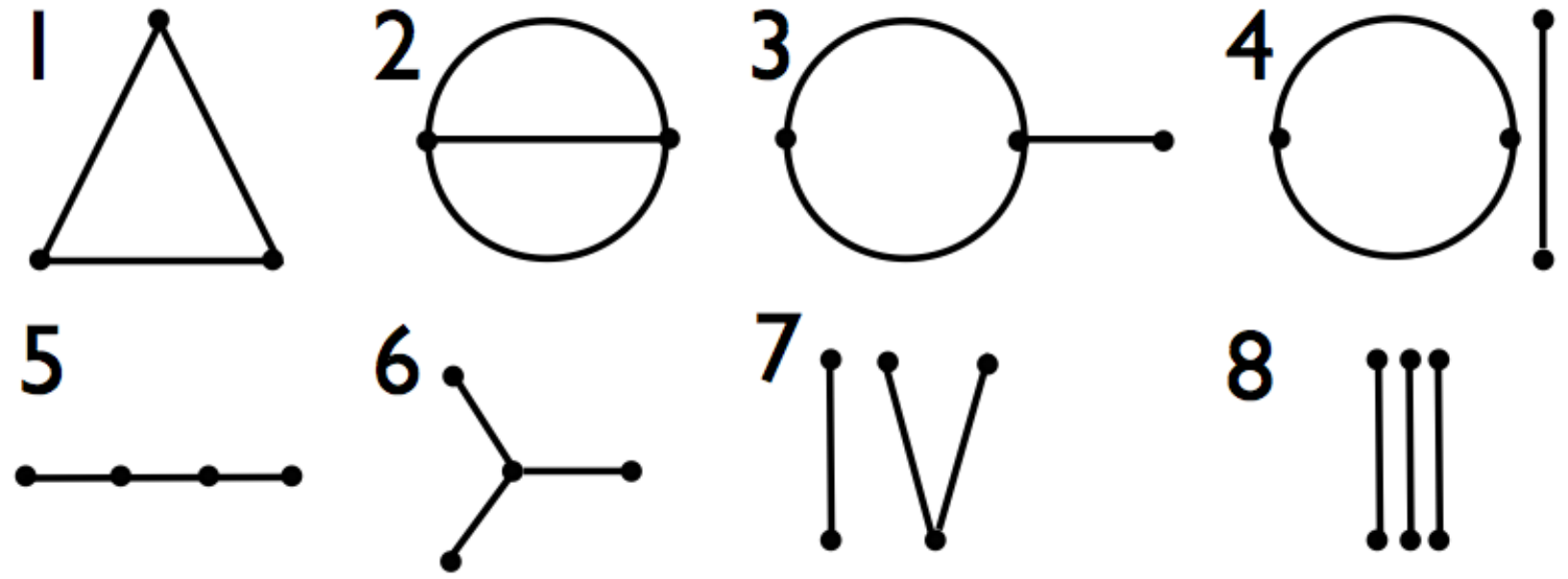,width=0.49\textwidth}
\caption{Diagrams corresponding to the cubic cumulants $\W$.}
\label{C3}
\end{center}
\end{figure}
We have thus only three possible values of $G$, 
\beq
G_{ab,ab}=G_1\, , \ \ G_{ab,ac}=G_2\, , \ \  G_{ab,cd}=G_3 \ ,
\eeq
and eight possible values of $\mathcal W$ as pictorially listed in fig. \ref{C3}:
\beq
\W_{ab,bc,ca}=\W_1\,,\ \ \W_{ab,ab,ab}=\W_2\,,\ \ \W_{ab,ab,ac}=\W_3\,,\ \ \W_{ab,ab,cd}=\W_4\,,\ \  
\eeq
\beq
\W_{ab,ac,bd}=\W_5\,,\ \ \W_{ab,ac,ad}=\W_6\,,\ \ \W_{ac,bc,de}=\W_7\,,\ \ \W_{ab,cd,ef}=\W_8\,,\ \ 
\eeq
We want to recast the cubic part of the free energy in the following form: 
\beqa
 \sum_{(ab),(cd),(ef)}\W_{ab,cd,ef}\lambda_{ab}\lambda_{cd}\lambda_{ef} & = &  \omega_1 \sum_{abc}\lambda_{ab}\lambda_{bc}\lambda_{ca}+\omega_2 \sum_{ab}\lambda_{ab}^3+        
\nonumber
\\
& +& \omega_3 \sum_{abc}\lambda_{ab}^2\lambda_{ac}+\omega_4 \sum_{abcd}\lambda_{ab}^2\lambda_{cd}+\omega_5 \sum_{abcd}\lambda_{ab}\lambda_{ac}\lambda_{bd}+
\nonumber
\\
& + &  \omega_6 \sum_{abcd}\lambda_{ab}\lambda_{ac}\lambda_{ad}+\omega_7 \sum_{abcde}\lambda_{ac}\lambda_{bc}\lambda_{de}+\omega_8 \sum_{abcdef}\lambda_{ab}\lambda_{cd}\lambda_{ef}                      \, .
\label{cubicf}
\eeqa
The above identity leads to the following relationships between the $\omega$'s and the $\W$'s \cite{temesvari2002generic}:
\beqa
\omega_1 & = & \W_1 -3 \W_5+3 \W_7-\W_8
\\
\omega_2 & = & {1 \over 2}\W_2-3 \W_3+{3 \over 2}\W_4+3 \W_5+2 \W_6 -6 \W_7+2 \W_8 
\\
\omega_3 & = &  3 \W_3 - 3 \W_4 - 6 \W_5 - 3 \W_6 + 15 \W_7 - 6 \W_8
\\
\omega_4 & = & {3 \over 4}(\W_4-2 \W_7+\W_8)
\\
\omega_5 & = & 3\W_5-6\W_7+3\W_8
\\
\omega_6 & = & \W_6-3\W_7+2\W_8
\\
\omega_7 & = & {3 \over 2}\W_7-{3 \over 2}\W_8
\\
\omega_8 & = & {1 \over 8}\W_8
\eeqa
From the definition (\ref{FDEF}) we easily see that  the coefficients of $F(\lambda)$ can be related to spin averages, in particular $G$ is precisely the dressed propagator:
\beq
G_{(ab),(cd)}\equiv -{\partial^2 \over \partial \lambda_{ab}\partial \lambda_{cd}}F(\lambda)=N\langle \delta \Q_{ab}\delta \Q_{cd}\rangle
\label{DL2}
\eeq
In the following and in the previous expression averages are always computed at $\lambda_{ab}=0$.
Assuming that we are in a replica symmetric  phase we obtain that $G_{(ab),(cd)}$ can take three possible values depending on whether there are two, three or four different replica indexes. The corresponding values are:
\beqa
G_1 & \equiv & N \langle \delta \Q_{12}^2\rangle = {1 \over N}  \sum_{ij}(\overline{\langle s_i s_j\rangle^2}-q^2) 
\label{G1}
\\
G_2 & \equiv & N \langle \delta \Q_{12} \delta \Q_{13}\rangle = {1 \over N}\sum_{ij}(\overline{\langle s_i s_j\rangle \langle{s_i}\rangle \langle{s_j}\rangle}-q^2) 
\label{G2}
\\
G_3 & \equiv & N \langle \delta \Q_{12} \delta \Q_{34}\rangle = {1 \over N}  \sum_{ij}(\overline{ \langle{s_i}\rangle^2 \langle{s_j}\rangle^2}-q^2)
\label{G3}
\eeqa
The cubic terms are given by the third derivative:
\beq
\W_{(ab),(cd),(ef)}\equiv -{\partial^3 \over \partial \lambda_{ab}\partial \lambda_{cd}\partial \lambda_{ef}}F(\lambda)=N^2\langle \delta \Q_{ab}\delta \Q_{cd}\delta \Q_{ef} \rangle_c =N^2 \langle \delta \Q_{ab}\delta \Q_{cd}\delta \Q_{ef} \rangle 
\label{DL3}
\eeq
where the suffix $c$ stands for connected functions with respect to the overlaps (not with respect to the spins) and the second equality follows from the fact that the average of $\delta \Q_{ab}$ is zero by definition.
The cubic cumulants can take eight possible values:
\beqa
\W_1 & = &N^2 \langle \delta \Q_{12}\delta \Q_{23}\delta \Q_{31} \rangle =
\nonumber
\\
& = & {1 \over N}  \sum_{ijk}\overline{\langle s_i s_j \rangle \langle s_j s_k \rangle \langle s_k s_i \rangle} - {3 q}  \sum_{ij }\overline{\langle s_i s_j\rangle \langle{s_i}\rangle \langle{s_j}\rangle}+ 2 N^2 q^3 
\label{W1} 
\\
\W_2 & = & {N^2} \langle \delta \Q_{12}^3  \rangle =
\nonumber
\\
& = & {1 \over  N}  \sum_{ijk}\overline{\langle s_i s_j s_k\rangle^2 } - {3 q }  \sum_{ij }\overline{\langle s_i s_j\rangle^2}+  2 N^2 q^3 
\label{W2}
\\
\W_3 & = &{N^2} \langle \delta \Q^2_{12}\delta \Q_{13}  \rangle =
\nonumber
\\
& = & {1 \over  N}  \sum_{ijk}\overline{\langle s_i s_j s_k\rangle \langle s_i s_j \rangle \langle s_k \rangle } - {2 q }  \sum_{ij }\overline{\langle s_i s_j\rangle \langle s_i \rangle \langle s_j \rangle}- { q }  \sum_{ij }\overline{\langle s_i s_j\rangle^2}+  2 N^2 q^3 
\label{W3}
\\
\W_4 & = &{N^2} \langle \delta \Q_{12}^2\delta \Q_{34}  \rangle =
\nonumber
\\
& = & {1 \over  N}  \sum_{ijk}\overline{\langle s_i s_j\rangle^2 \langle s_k \rangle^2 } - {2 q }  \sum_{ij }\overline{\langle s_i \rangle^2 \langle s_j \rangle^2 }- { q }  \sum_{ij }\overline{\langle s_i s_j\rangle^2}+  2 N^2 q^3 
\label{W4}
\\
\W_5 & = & {N^2} \langle \delta \Q_{12}\delta \Q_{13}\delta \Q_{24}  \rangle =
\nonumber
\\
& = & {1 \over  N}  \sum_{ijk}\overline{\langle s_i s_j\rangle \langle s_i s_k\rangle \langle s_k \rangle \langle s_j \rangle } - {2 q }  \sum_{ij }\overline{\langle s_i s_j\rangle \langle s_i \rangle \langle s_j \rangle }- { q }  \sum_{ij }\overline{\langle s_i \rangle^2 \langle s_j\rangle^2}+  2 N^2 q^3 
\label{W5}
\\
\W_6 & = & {N^2} \langle \delta \Q_{12}\delta \Q_{13}\delta \Q_{14}  \rangle =
\nonumber
\\
& = & {1 \over  N}  \sum_{ijk}\overline{\langle s_i s_j s_k\rangle \langle s_i\rangle \langle s_j \rangle \langle s_k \rangle } - {3 q }  \sum_{ij }\overline{\langle s_i s_j\rangle \langle s_i \rangle \langle s_j \rangle }+  2 N^2 q^3 
\label{W6}
\\
\W_7 & = & {N^2} \langle \delta \Q_{12}\delta \Q_{13}\delta \Q_{45}  \rangle =
\nonumber
\\
& = & {1 \over  N}  \sum_{ijk}\overline{\langle s_i s_j \rangle \langle s_k\rangle^2 \langle s_i\rangle \langle s_j \rangle } - {2 q }  \sum_{ij }\overline{\langle s_i\rangle^2 \langle s_j\rangle^2 }- {q }  \sum_{ij }\overline{\langle s_i s_j\rangle \langle s_i \rangle \langle s_j \rangle }+  2 N^2 q^3 
\label{W7}
\\
\W_8 & = &{N^2} \langle \delta \Q_{12}\delta \Q_{34}\delta \Q_{56}  \rangle =
\nonumber
\\
& = & {1 \over  N}  \sum_{ijk}\overline{\langle s_i \rangle^2 \langle s_j \rangle^2 \langle s_k\rangle^2 } - {3 q }  \sum_{ij }\overline{\langle s_i\rangle^2 \langle s_j\rangle^2 }+  2 N^2 q^3 
\label{W8}
\eeqa
Substituting the above expressions in the relationship between the $\omega$'s and the $\W$ we obtain:
\beqa
\omega_1 & = & {1 \over N} \sum_{ijk}\overline{\langle s_i s_j\rangle_c \langle s_j s_k\rangle_c \langle s_k s_i\rangle_c}
\label{omega1}
\\
\omega_2 & = & {1 \over 2 N} \sum_{ijk}\overline{\langle s_i s_j s_k\rangle_c^2 }
\label{omega2}
\\
\omega_3 & = &{3 \over  N}  \sum_{ijk}\overline{\langle s_i s_j s_k\rangle_c \langle s_i s_j \rangle_c \langle s_k \rangle }
\label{omega3}
\\
\omega_4 & = & {3 \over 4 N}  \sum_{ijk}\left[\overline{\langle s_i s_j\rangle_c^2 \langle s_k \rangle^2 }-\overline{\langle s_i s_j\rangle_c^2}\ \overline{ \langle s_k \rangle^2 }\right]
\label{omega4}
\\
\omega_5 & = & {3 \over  N}  \sum_{ijk}\overline{\langle s_i s_j\rangle_c \langle s_i s_k\rangle_c \langle s_k \rangle \langle s_j \rangle }
\label{omega5}
\\
\omega_6 & =& {1 \over  N}  \sum_{ijk}\overline{\langle s_i s_j s_k\rangle_c \langle s_i\rangle \langle s_j \rangle \langle s_k \rangle }
\label{omega6}
\\
\omega_7 & = & {3 \over 2 N}\sum_{ijk}\left[\overline{\langle s_i s_j \rangle_c\langle s_i\rangle \langle s_j\rangle \langle s_k\rangle^2}-\overline{\langle s_i s_j \rangle_c\langle s_i\rangle \langle s_j\rangle}\ \overline{\langle s_k\rangle^2}\right]
\label{omega7}
\\
\omega_8 & = & {N^2 \over 8}\overline{\left( q_J-q\right)^3}
\label{omega8}
\eeqa
We note that upon passing from the $\W$'s to the $\omega$'s there is increase in symmetry and simplicity, in particular we see that due to various cancellations $\omega_1$,  $\omega_2$,  $\omega_3$,  $\omega_5$ and $\omega_6$ have a single disorder average, $\omega_4$ and $\omega_7$ have two disorder average and only $\omega_8$ has three disorder averages.

We now turn to the fourth order contribution  that involves the $23$ diagrams shown in Fig. (\ref{C4}). These same diagrams have been also studied by Temesvari (see Appendix A in \cite{temesvari2007replica}, note that we use a different name convention).
\begin{figure}[t!]
\begin{center}
\epsfig{file=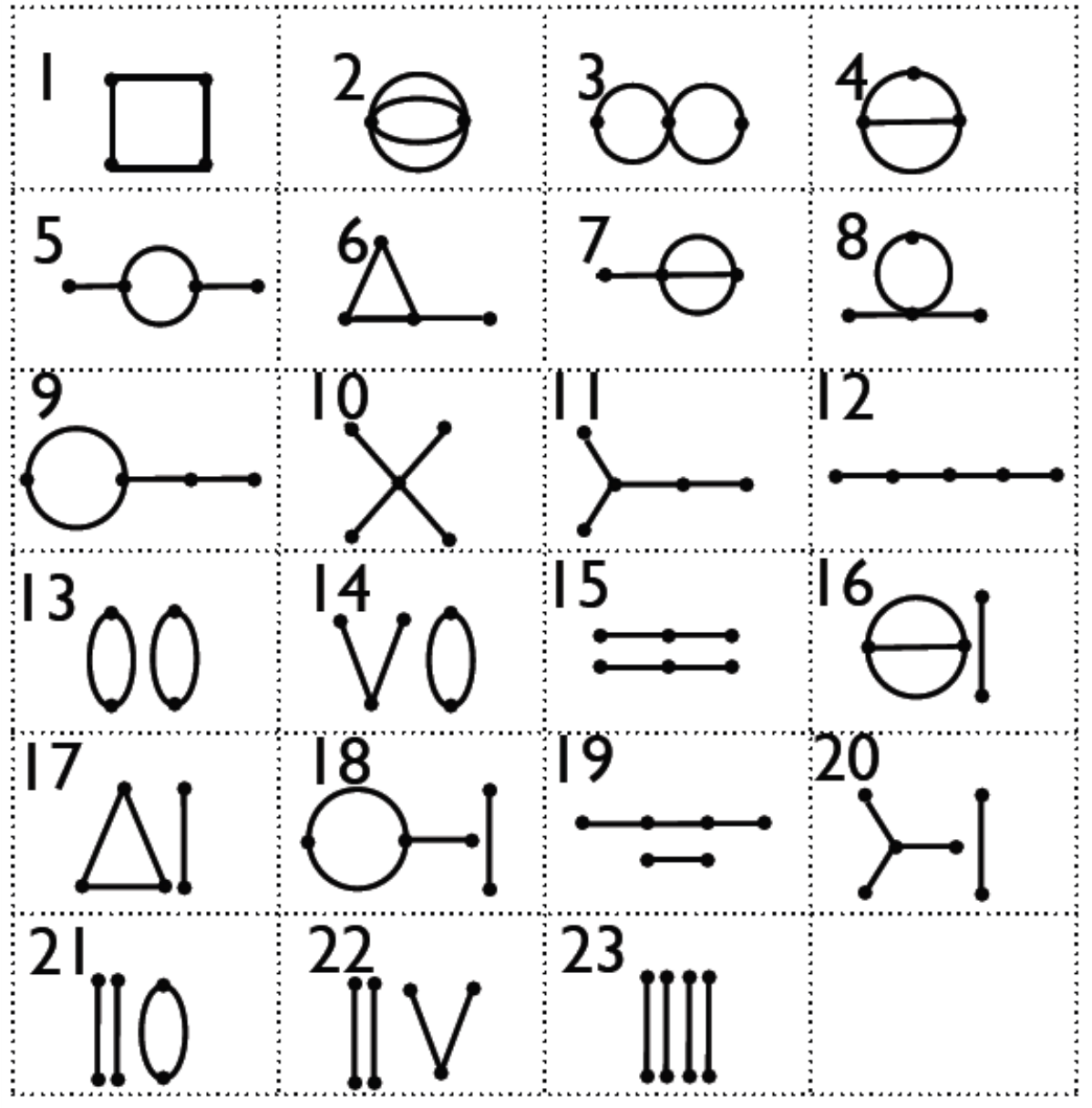,width=0.49\textwidth}
\caption{Diagrams corresponding to the quartic cumulants $\mathcal Y$.}
\label{C4}
\end{center}
\end{figure}
The expressions of the cumulants $\Y_{ab,cd,ef,gh}$ in terms of the physical observables can be obtained by differentiation as we did before for the third order, 
cf. Eq. (\ref{W1}-\ref{W8}). 
Again, we are not interested  directly in the $\Y$ cumulants but, rather, in those linear combinations of theirs corresponding to the unrestricted sums over replicas indexes. In other words,  we want to determine the $23$ connected correlation functions $\y_i$ that satisfy the following equation:
\beq
{1\over 24} \sum_{(ab),(cd),(ef),(gh)}\Y_{ab,cd,ef,gh}\lambda_{ab}\lambda_{cd}\lambda_{ef}\lambda_{gh}={1 \over 24}\left[\y_1 \Tr \lambda^4 + \y_2 \sum_{ab}\lambda_{ab}^4+\y_3 \sum_{abc}\lambda_{ab}^2\lambda_{ac}^2 + \y_4 \sum_{abc} \lambda_{ab}^2\lambda_{ac}\lambda_{cb}+\dots\right]
\label{quarcon}
\eeq
Inasmuch as in the cubic case, we should first associate to the coefficients $\Y_i$ the appropriate averages of the overlap (corresponding to Eqs. (\ref{W1}-\ref{W8}) for the third order) and then separately determine the connection between the $\Y_i$'s
 and the $\y_i$. Both computations are reported in Appendix \ref{App:4th}.
The results can, then, be used to derive the analog of expressions (\ref{omega1}-\ref{omega8}). In spite of the complexity of the intermediate passages it turns out that the result is particularly simple for the four terms explicitly included in Eq. (\ref{f:Gibbs_fe}), only three of which are those relevant to compute the $\mu$ coefficient of the logarithmic decay. We find:
\beqa
\y_1 & = & {3 \over N} \sum_{ijkl}\overline{\langle s_i s_j\rangle_c \langle s_j s_k\rangle_c \langle s_k s_l\rangle_c\langle s_l s_i\rangle_c}
\label{yy1}
\\
\y_2 & = & {1 \over 2 N} \sum_{ijkl}\overline{\langle s_i s_j s_k s_l\rangle_c^2 }
\label{yy2}
\\
\y_3 & = &{3 \over  N}  \sum_{ijkl}\overline{\langle s_i s_j s_k s_l\rangle_c \langle s_i s_j \rangle_c \langle s_k s_l \rangle_c }
\label{yy3}
\\
\y_4 & = & {6 \over  N}  \sum_{ijk}\overline{\langle s_i s_j s_k\rangle_c \langle s_i s_j s_l \rangle_c \langle s_l s_k \rangle_c } \ .
\label{yy4}
\eeqa
We are now in position to discuss why we retained only two cubic diagrams and four quartic diagrams in the Gibbs free energy (\ref{f:Gibbs_fe}).
The key point is that the dynamical correlation $\delta Q(1,2)$, for all the three transitions outlined in sec. \ref{Outline}, for $n=0,1$, satisfies the relationship
\beq
\int d1 \, Q(1,2) = \delta C(0) + (n-1) \delta C(\infty) = 0 
\label{int1}
\eeq
because $\delta C(0)=\delta C(\infty)=0$.
More generically, one can argue that any object formed  from $\delta Q(1,2)$ by means of products and index integrations, 
that depends only on {\it one} index, {\it e.g.} 
\beq
A(1) \equiv \int [\delta Q(1,2)]^5\ [\delta Q(2,3)]^6\ \ \delta Q(2,4) \ \delta Q(4,1) \ d2 \ d3\  d4 \ ,
\eeq
{\it vanishes} because, upon computing it, one ends up with an expression that only depends  on powers of $\delta C(0)$ and $\delta C(\infty)$. Actually, this is the same expression that one would obtain plugging into the above expression a replica symmetric matrix with diagonal elements equal to $\delta C(0)$ and off-diagonal elements equal to $\delta C(\infty)$.

The same is naturally true for objects that depend on no index at all. Therefore, we can neglect all {\it disconnected diagrams} in the Gibbs free energy as they lead to terms containing factors that do not depend on either $1$ or $2$  in the equation (\ref{eq:Gibbs_verteces}) obtained by differentiation with respect to $\delta Q(1,2)$.

For the same reason, we can also neglect {\it diagrams with dangling hands} in the Gibbs free energy. 
Indeed, they contribute two types of terms to the equation: either a term with a dangling hand (that vanishes because of Eq. (\ref{int1})) or terms that depends only on index $1$ or $2$ separately and, thus, vanish. 
More generically, any diagram that can be  disconnected removing {\it one vertex} yields a vanishing contribution.  

Diagrams with dangling hands are the simplest diagrams satisfying this property but not the only ones. 
Indeed, consistently, we could have also ignored the term proportional to $y_3$ in expression (\ref{f:Gibbs_fe}) from the beginning, since removing the central vertex in the third diagram in Fig. \ref{C4} we have two disconnected graphs.

\section{Inversion of the Legendre 
Transform: relations between cumulants and vertex coefficients.}
\label{Sec:VertexCumul}

In this section we express the coefficients of the Gibbs free energy in terms of those of the free energy obtained in the previous section.
The Gibbs Free energy is defined as the Legendre transform of the Free energy $F(\lambda)$:
\beq
G(\delta Q) \equiv F(\lambda)+\sum_{(ab)}\lambda_{ab}\delta Q_{ab}
\eeq
where $\lambda$ is a function of $\delta Q_{ab}$ according to the following implicit equation:
\beq
\delta Q_{ab}=-{\partial F \over \partial \lambda_{ab}} .
\eeq
On the other hand, the free energy $F$ is the Legendre transform of the Gibbs free energy $G$ with:
\beq
\lambda_{ab}={\partial G \over \partial \delta Q_{ab}} \ .
\eeq
We consider the free energy expansion Eq. (\ref{eq:FEN4th}), taking into account only those terms eventually relevant to describe the critical slowing down:
\begin{eqnarray}
F(\lambda)&=&-{1 \over 2}\sum_{(ab),(cd)}\lambda_{ab}G_{ab,cd}\lambda_{cd}\\
\nonumber
&&-\frac{\omega_1}{ 6}\omega_1\Tr \lambda^3 -\frac{\omega_2}{6}\sum_{ab}\lambda_{ab}^3\\
\nonumber
&&-\frac{\y_1}{24}\Tr \lambda^4-\frac{\y_2}{24}\sum_{ab}\lambda^4_{ab}
\\
\nonumber
&&-\frac{\y_4}{24}\sum_{abc}\lambda_{ab}^2\lambda_{ac}
\lambda_{ef}
\end{eqnarray}
leading to 
\begin{eqnarray}
\label{eq:deQ_la}
\delta Q_{ab}&=&-\frac{\partial F}{\partial \lambda_{ab}} = G_{(ab)(cd)}\lambda_{cd}
\\
&&+\omega_1(\lambda^2)_{ab}+\omega_2\lambda_{ab}^2
\nonumber
\\
\nonumber
&&
+\frac{\y_1}{3}(\lambda^3)_{ab}+\frac{\y_2}{3}\lambda_{ab}^3
\\
\nonumber
&&+\frac{\y_4}{6}\lambda_{ab}(\lambda^2)_{ab}+\frac{\y_4}{12}
\sum_c\left(
\lambda_{ac}^2\lambda_{cb}
+\lambda^2_{bc}\lambda_{ca}
\right)
\end{eqnarray}

where
\begin{equation}
G_{(ab)(cd)}=\left\{\begin{array}{ll} 
G_1\quad & (ab)=(cd)
\\ G_2\quad & (b=d)\vee (a=c) 
\\ G_3 \quad& a\neq c \wedge b\neq d
\end{array}\right. 
\end{equation}
The inverse $M=G^{-1}$ matrix
displays the generic form
\begin{eqnarray}
M_{(ab)(cd)}
&=& 
r 
\left(\delta_{ac}\delta_{bd}+\delta_{ad}\delta_{bc}\right)
\\
\nonumber
&&+(M_2-M_3) \left(\delta_{ac}+\delta_{ad}+\delta_{bc}+\delta_{bd}\right)\\
\nonumber
&&+M_3
\end{eqnarray}
with
\begin{equation}
r\equiv M_1-2M_2+M_3=\frac{1}{G_1-2G_2+G_3}
\end{equation}

Using the above properties and neglecting terms without both indexes $a$ and $b$ (irrelevant in the present context) we can invert Eq. (\ref{eq:deQ_la}) yielding, to the fourth order
\begin{eqnarray}
\lambda_{ab}&=&r \dQ_{ab}-\omega_1r^3\left(\dQ^2\right)_{ab}
-\omega_2r^3\dQ_{ab}^3\\
\nonumber
&&+\left(2 r^5\omega_1^2-r^4\frac{\y_1}{3}\right)\left(\dQ^3\right)_{ab}
\\
\nonumber
&&+\left(2 r^5\omega_2^2-r^4\frac{\y_2}{3}\right)\dQ_{ab}^3
\\
\nonumber
&&+\left( r^5\omega_1\omega_2-r^4\frac{\y_4}{12}\right)\sum_c\left(\dQ_{ac}^2 \dQ_{cb}+\dQ_{ac}\dQ_{cb}^2\right)
\\
\nonumber
&&+\left(2 r^5\omega_1\omega_2-r^4\frac{\y_4}{6}\right)\dQ_{ab} \left(\dQ^2\right)_{ab}
\end{eqnarray}
Comparing the above expression with Eqs. (\ref{eq:Gexp}), (\ref{eq:Gibbs_verteces})
we find the relationships between the cumulants $\omega$, $\y$ and  the vertex coefficients $w$,$y$
\begin{eqnarray}
y_1&=&-6r^5\omega_1^2+r^4\y_1
\label{eq:y1_r}
\\
y_2&=&-6r^5\omega_2^2+r^4\y_2
\label{eq:y2_r}
\\
y_4&=&-12r^5\omega_1\omega_2+r^4\y_4
\label{eq:y4_r}
\\
w_1&=&r^3\omega_1
\label{eq:w1_r}
\\
w_2&=&r^3\omega_2
\label{eq:w2_r}
\end{eqnarray}

At the tricritical point, Tthe expression for the logarithmic decay parameter $\mu$, cf. Eq. (\ref{muin1}) can, thus, be expressed in terms of cumulants as 
\begin{equation}
\mu=-\frac{y_1+y_2-y_4}{3 w_1}=-r\frac{\y_1+\y_2-\y_4}{3 \omega_1}\, , \ \ \ (\lambda=1)
\end{equation}
where we explicitly used the fact that the exponent parameter is $\lambda=1=\omega_2/\omega_1$.
From Eqs. (\ref{eq:y1_r})-(\ref{eq:w2_r})
we notice that though each vertex coefficient singularly diverges as $r^{-5}$, their combination $y_1+y_2-y_4$, for $w_1=w_2$,
diverges as $r^{-4}$, thus yielding a finite $\mu$ when power-law critical slowing down (described by means of the third order expansion when $\lambda \neq 1$)
is no longer defined.
To clearly see this we can express the quartic susceptibilities in term of the coupling constants:
\beq
\y_1+\y_2-\y_4 = 6\frac{(w_1-w_2)^2}{r^5} + \frac{y_1+y_2-y_4}{r^4}
\eeq
This shows that whenever $w_1 \neq w_2$ the critical behaviour of the quartic susceptibility is $O(r^{-5})$ and it is controlled by the cubic coupling constants. Instead, when $w_1=w_2$ the quartic susceptibility is less divergent $O(r^{-4})$ and it is controlled by the quartic coupling constants.

\section{Conclusions}
\label{conclusions}

In this paper we have demonstrated that the structure of the replicated Gibbs free energy near a critical point characterized by $w_1=w_2$ implies a logarithmic decay of dynamical correlations. This allows to characterize the asymptotic critical dynamics in a variety of systems where the equilibrium statics can be studied by means of the replica method but the
microscopic dynamical equations are difficult to be solved, including Ising spin-glass models, Potts spin-glass models and hard-spheres model in the limit of infinite dimension.

The connection between static and dynamics is also quantitative, in the sense that the parameter controlling the logarithmic decay can be read from the static Gibbs free energy and, thus, it can be expressed 
in terms of connected correlation functions  of the overlap fluctuations that can be measured statically from equilibrium configurations. This is significant from the point of view of numerical simulations of glassy systems, as often one can use clever algorithms to obtain equilibrium configurations much faster than the standard dynamical microscopic evolution \cite{marinari1992simulated,hukushima1996exchange,ninarello2017models}.

The emergence of logarithmic slowing down, being a consequence solely of the Gibbs free energy structure, has a great deal of universality. Indeed many models, that can be utterly different from each other at the microscopic level, can in principle be described by the same Landau theory. Note that we have written the expression of $\mu$ in terms of observables for spin systems but it can be easily rewritten for particles systems as we explained in section \ref{Outline}. Thus the relationship between $\mu$ and experimental observables is completely general: it would be important to work out in full the connection between these cumulants and higher-order non-linear susceptibilities that can be measured in experiments \cite{albert2019third}.

It is also interesting to mention two instances in which there is instead no connection between logarithmic slowing down and connected correlation functions of the overlap fluctuations. This is provided by  Fredrikson-Andersen Kinetically constrained model on the Bethe lattice with either random pinning \cite{ikeda2017fredrickson} or with non-homogeneous facilitation \cite{sellitto2010dynamic,arenzon2012microscopic}: the analytical solution of these models \cite{perrupato2022exact} has indeed allowed to demonstrate and quantify the logarithmic slowing down as given by eq. (\ref{clog}) but a thermodynamic analysis of the model has revealed that there is no connection between the observed MCT-like dynamics and connected correlation functions of the overlap that are not divergent at the critical point \cite{perrupato2023thermodynamics}, {\it i.e.} eq. (\ref{muin2}) does not hold.

The results presented here have a mean-field nature and  their relevance for realistic models in finite dimension, say two and three, in principle is not granted. This issue notwithstanding,  analytical predictions like those derived here and more generally those of Mode-Coupling-Theory are typically used to describe successfully numerical and experimental data in the context of supercooled liquids and colloids \cite{Goetze09,Sjogren91,Zaccarelli02,Sciortino03,Sciortino03b}. A relevant exception is the discontinuous dynamical arrest transition that is known to become a cross-over in finite dimension: it has been argued that this phenomenon is precisely due to long-wavelength fluctuations that destroy the mean-field theory below the upper critical dimension eight \cite{franz2011field,rizzo2014long,rizzo2016dynamical,rizzo2015qualitative,rizzo2020solvable,elizondo2022subaging}.
In the Spin-Glass literature the relevance of mean-field theory for realistic systems is an essential question \cite{bray1980renormalisation,lubensky2023renormalization,angelini2022unexpected} but  it is typically discussed in a purely static context and there is at present no understanding of the fate of the dynamical power-law (and logarithmic) decays found in mean-field theory.

It should be also noted that we have been considering equilibrium dynamics, therefore the results are limited to the temperature regime in which the system can still be thermalized and there is no aging. Nonetheless it is not unlikely that, with more effort, the analysis can be extended to the aging regime, as it is known that static replicated theories can be connected to off-equilibrium dynamics as well \cite{cugliandolo1993analytical,caltagirone12c}.

As a final technical remark we note that upon passing from restricted to unrestricted replica summations the corresponding coefficients considerably simplify.  This can be seen comparing the coefficients $\mathcal{W}_i$ in the free energy expansion (\ref{eq:FEN4th}) with the $\omega_i$ in Eq. (\ref{cubicf}) and comparing the $\mathcal{Y}_i$ with the $\y_i$ in Eq. (\ref{quarcon}). It turns out that one can find a simple set of diagrammatic rules to directly compute  the unrestricted coefficients, {\it i.e.}, the $\omega_i$, the $\y_i$ and those at higher orders, without the lengthy intermediate passages \footnote{L. Leuzzi and T. Rizzo, to be published}.

\appendix

\section{Fourth order cumulants}
\label{App:4th}
The fourth order coefficients of the Free energy read:
\begin{eqnarray}
\Y_{1} &=&N^3\langle\delta Q_{ab} \delta Q_{bc}  \delta Q_{cd}  \delta Q_{da}  \rangle_c 
\\
\nonumber
&=& \frac{1}{N}\sum_{ijkl}{\overline{\langle s_is_j \rangle\langle s_js_k \rangle\langle s_ks_l \rangle\langle s_ls_i \rangle}}
 -4q \sum_{ijk}{\overline{\langle s_i \rangle\langle s_is_j \rangle\langle s_js_k \rangle\langle s_k \rangle}}
 \\
 \nonumber
 &&
 +2 Nq^2\sum_{ij}{\overline{\langle  s_i\rangle^2\langle  s_j\rangle^2}}
+4Nq^2\sum_{ij}{\overline{\langle  s_i\rangle\langle  s_i s_j\rangle\langle  s_j\rangle}}
 - 3q^4N^3-2 NG_2^2-NG_3^2
\label{Y1}
\\
\Y_{2} &=&N^3\langle\delta Q_{12}^4  \rangle_c
\\
\nonumber
&=& \frac{1}{N}\sum_{ijkl}{\overline{\langle  s_is_js_ks_l\rangle^2}}
-4q\sum_{ijk}{\overline{\langle  s_is_js_k\rangle^2}}
 +6Nq^2\sum_{ij}{\overline{\langle  s_i s_j\rangle^2}}
 - 3q^4N^3-3NG_1^2
\\
\Y_{3} &=&N^3\langle\delta Q_{ab}^2 \delta Q_{bc}^2  \rangle_c
\\
\nonumber
&=& \frac{1}{N}\sum_{ijkl}{\overline{\langle s_is_j \rangle\langle s_is_js_ks_l \rangle\langle s_ks_l \rangle}}
-4q \sum_{ijk}{\overline{\langle s_is_j  \rangle\langle s_is_js_k  \rangle\langle s_k \rangle}}
\\
\nonumber
&&
+4Nq^2\sum_{ij}{\overline{\langle  s_i\rangle\langle  s_i s_j\rangle\langle  s_j\rangle}}
 +2Nq^2\sum_{ij}{\overline{\langle  s_i s_j\rangle^2}}
 - 3q^4N^3-NG_1^2-2NG_2^2
\\
\Y_{4} &=&N^3\langle\delta Q_{ab}\delta Q_{bc} \delta Q_{ac}^2  \rangle_c
\\
\nonumber
&=& \frac{1}{N}\sum_{ijkl}{\overline{\langle s_is_js_k \rangle\langle s_is_js_l \rangle\langle s_ks_l \rangle}}
 -2q \sum_{ijk}{\overline{\langle s_is_j  \rangle\langle s_is_js_k  \rangle\langle s_k \rangle}}
 -2q\sum_{ijk}{\overline{\langle s_is_j \rangle \langle s_js_k \rangle \langle s_k s_i\rangle}}
 \\
 \nonumber
 &&
+5Nq^2\sum_{ij}{\overline{\langle  s_i\rangle\langle  s_i s_j\rangle\langle  s_j\rangle}}
+ Nq^2\sum_{ij}{\overline{\langle  s_i s_j\rangle^2}}
 - 3q^4N^3-NG_1G_2-2NG_2^2
\\
\Y_{5} &=&N^3\langle\delta Q_{ab} \delta Q_{bc}^2 \delta Q_{cd} \rangle_c
\\
\nonumber
&=& \frac{1}{N}\sum_{ijkl}{\overline{\langle s_i \rangle\langle s_is_js_k  \rangle\langle  s_js_ks_l \rangle\langle s_l \rangle}}
 -2q \sum_{ijk}{\overline{\langle s_is_j  \rangle\langle s_is_js_k  \rangle\langle s_k \rangle}}
 -2q \sum_{ijk}{\overline{\langle s_i \rangle\langle s_is_j \rangle\langle s_js_k \rangle\langle s_k \rangle}}
 \\
 \nonumber
 &&
+ Nq^2\sum_{ij}{\overline{\langle  s_i\rangle^2\langle  s_j\rangle^2}}
 +4Nq^2\sum_{ij}{\overline{\langle  s_i\rangle\langle  s_i s_j\rangle\langle  s_j\rangle}}
 +Nq^2\sum_{ij}{\overline{\langle  s_i s_j\rangle^2}}
 - 3q^4N^3- 2NG_2^2-NG_1G_3
\\
\Y_{6} &=&N^3\langle\delta Q_{ab} \delta Q_{bc} \delta Q_{ca} \delta Q_{cd}  \rangle_c
\\
\nonumber
&=& \frac{1}{N}\sum_{ijkl}{\overline{\langle  s_is_j\rangle\langle s_is_ks_l  \rangle\langle s_ks_i \rangle\langle s_l \rangle}}
 -  q\sum_{ijk}{\overline{\langle s_i s_js_k\rangle\langle s_i \rangle\langle s_j \rangle\langle s_k \rangle}} 
 \\
 \nonumber
 &&
  -2 q\sum_{ijk}{\overline{\langle s_i \rangle\langle s_is_j \rangle\langle s_js_k \rangle\langle s_k \rangle}}
 -q\sum_{ijk}{\overline{\langle s_is_j \rangle \langle s_js_k \rangle \langle s_k s_i\rangle}}
 \\
 \nonumber
 &&
+  Nq^2\sum_{ij}{\overline{\langle  s_i\rangle^2\langle  s_j\rangle^2}}
+5 Nq^2\sum_{ij}{\overline{\langle  s_i\rangle\langle  s_i s_j\rangle\langle  s_j\rangle}}
 - 3q^4N^3 -2NG_2^2-NG_2G_3
\\
\Y_{7} &=&N^3\langle\delta Q_{ab} \delta Q_{bc}^3  \rangle_c
\\
\nonumber
&=& \frac{1}{N}\sum_{ijkl}{\overline{\langle s_i  s_js_ks_l\rangle\langle s_i  \rangle\langle s_js_ks_l \rangle}}
-q\sum_{ijk}{\overline{\langle  s_is_js_k\rangle^2}}
 -3q \sum_{ijk}{\overline{\langle s_is_j  \rangle\langle s_is_js_k  \rangle\langle s_k \rangle}}
 \\
 \nonumber
 &&
 +3Nq^2\sum_{ij}{\overline{\langle  s_i\rangle\langle  s_i s_j\rangle\langle  s_j\rangle}}
 +3Nq^2\sum_{ij}{\overline{\langle  s_i s_j\rangle^2}}
 - 3q^4N^3-3NG_1G_2
\\
\Y_{8} &=&N^3\langle\delta Q_{ab} \delta Q_{bc}^2\delta Q_{bd}  \rangle_c
\\
\nonumber
&=& \frac{1}{N}\sum_{ijkl}{\overline{\langle s_is_js_ks_l  \rangle\langle  s_i\rangle\langle  s_js_k\rangle\langle s_l \rangle}}
 -2q\sum_{ijk}{\overline{\langle s_i s_js_k\rangle\langle s_i \rangle\langle s_j \rangle\langle s_k \rangle}} 
-2q  \sum_{ijk}{\overline{\langle s_is_j  \rangle\langle s_is_js_k  \rangle\langle s_k \rangle}}
\\
\nonumber
&&
 +5Nq^2\sum_{ij}{\overline{\langle  s_i\rangle\langle  s_i s_j\rangle\langle  s_j\rangle}}
 +Nq^2\sum_{ij}{\overline{\langle  s_i s_j\rangle^2}}
 - 3q^4N^3-2NG_2^2-NG_1G_2
\end{eqnarray}
\newpage
\begin{eqnarray}
\Y_{9} &=&N^3\langle\delta Q_{ab}^2 \delta Q _{bc} \delta Q_{cd}\rangle_c
\\
\nonumber
&=& \frac{1}{N}\sum_{ijkl}{\overline{\langle s_is_j  \rangle\langle s_is_js_k  \rangle\langle s_ks_l  \rangle\langle s_l  \rangle}}
 -2q \sum_{ijk}{\overline{\langle s_i \rangle\langle s_is_j \rangle\langle s_js_k \rangle\langle s_k \rangle}}
 \\
 \nonumber
 &&-q \sum_{ijk}{\overline{\langle s_is_j  \rangle^2\langle s_k \rangle^2}}
 -q\sum_{ijk}{\overline{\langle s_is_j  \rangle\langle s_is_js_k  \rangle\langle s_k \rangle}}
 \\
 \nonumber
 &&
+2 Nq^2\sum_{ij}{\overline{\langle  s_i\rangle^2\langle  s_j\rangle^2}}
+3 Nq^2\sum_{ij}{\overline{\langle  s_i\rangle\langle  s_i s_j\rangle\langle  s_j\rangle}}
+Nq^2\sum_{ij}{\overline{\langle  s_i s_j\rangle^2}}
 - 3q^4N^3-NG_1G_2 -2NG_2G_3
\\
\Y_{10} &=&N^3\langle\delta Q_{ab} \delta Q_{ac} \delta Q_{ad} \delta Q_{ae} \rangle_c
\\
\nonumber
&=& \frac{1}{N}\sum_{ijkl}{\overline{\langle s_is_js_ks_l \rangle\langle  s_i\rangle\langle s_j \rangle\langle s_k \rangle\langle  s_l\rangle}}
 -4q  \sum_{ijk}{\overline{\langle s_i s_js_k\rangle\langle s_i \rangle\langle s_j \rangle\langle s_k \rangle}} 
 \\
 \nonumber
 &&
+6Nq^2\sum_{ij}{\overline{\langle  s_i\rangle\langle  s_i s_j\rangle\langle  s_j\rangle}}
 - 3q^4N^3-3NG_2^2
\\
\Y_{11} &=&N^3\langle\delta Q_{ab} \delta Q_{bc} \delta Q_{bd} \delta Q_{de}  \rangle_c
\\
\nonumber
&=& \frac{1}{N}\sum_{ijkl}{\overline{\langle s_i  \rangle\langle s_is_js_k  \rangle\langle s_i  \rangle\langle s_ks_l  \rangle\langle s_l  \rangle}}
 -2q \sum_{ijk}{\overline{\langle s_i \rangle\langle s_is_j \rangle\langle s_js_k \rangle\langle s_k \rangle}}
  \\
 \nonumber
 &&
-q  \sum_{ijk}{\overline{\langle s_i \rangle^2\langle s_j \rangle\langle s_js_k \rangle\langle s_k \rangle}} 
 -q  \sum_{ijk}{\overline{\langle s_i s_js_k\rangle\langle s_i \rangle\langle s_j \rangle\langle s_k \rangle}} 
 \\
 \nonumber
 &&
 +2 Nq^2\sum_{ij}{\overline{\langle  s_i\rangle^2\langle  s_j\rangle^2}}
 +4Nq^2\sum_{ij}{\overline{\langle  s_i\rangle\langle  s_i s_j\rangle\langle  s_j\rangle}}
 - 3q^4N^3-NG_2^2-2NG_2G_3
\\
\Y_{12} &=&N^3\langle\delta Q_{ab} \delta Q_{bc}  \delta Q_{cd} \delta Q_{de}  \rangle_c
\\
\nonumber
&=& \frac{1}{N}\sum_{ijkl}{\overline{\langle  s_i\rangle\langle  s_is_j\rangle\langle  s_js_k\rangle\langle  s_ks_l\rangle\langle  s_l\rangle}}
 -2q  \sum_{ijk}{\overline{\langle s_i \rangle\langle s_is_j \rangle\langle s_js_k \rangle\langle s_k \rangle}}
\\
 \nonumber
 &&
 -2 q \sum_{ijk}{\overline{\langle s_i \rangle^2\langle s_j \rangle\langle s_js_k \rangle\langle s_k \rangle}} 
 +3Nq^2\sum_{ij}{\overline{\langle  s_i\rangle^2\langle  s_j\rangle^2}}
  \\
 \nonumber
 &&
+3Nq^2\sum_{ij}{\overline{\langle  s_i\rangle\langle  s_i s_j\rangle\langle  s_j\rangle}}
 - 3q^4N^3-NG_2^2-NG_3^2-NG_2G_3
\\
\Y_{13} &=&N^3\langle\delta Q_{ab}^2 \delta Q_{cd}^2  \rangle_c
\\
\nonumber
&=& \frac{1}{N}\sum_{ijkl}{\overline{\langle  s_is_j\rangle^2\langle  s_ks_l\rangle^2}}
 -4q \sum_{ijk}{\overline{\langle s_is_j  \rangle^2\langle s_k \rangle^2}}
 \\
 \nonumber
 &&
 +4Nq^2\sum_{ij}{\overline{\langle  s_i\rangle^2\langle  s_j\rangle^2}}
+2Nq^2\sum_{ij}{\overline{\langle  s_i s_j\rangle^2}}
 - 3q^4N^3-NG_1^2-2NG_3^2
\\
\Y_{14} &=&N^3\langle\delta Q_{ab}^2 \delta Q_{cd}  \delta Q_{de}\rangle_c
\\
\nonumber
&=& \frac{1}{N}\sum_{ijkl}{\overline{\langle  s_is_j\rangle^2\langle  s_k\rangle\langle  s_ks_l\rangle\langle  s_l\rangle}}
 -2q \sum_{ijk}{\overline{\langle s_i \rangle^2\langle s_j \rangle\langle s_js_k \rangle\langle s_k \rangle}} 
 -2q\sum_{ijk}{\overline{\langle s_is_j  \rangle^2\langle s_k \rangle^2}}
 \\
 \nonumber
 &&
 +4Nq^2\sum_{ij}{\overline{\langle  s_i\rangle^2\langle  s_j\rangle^2}}
+Nq^2\sum_{ij}{\overline{\langle  s_i\rangle\langle  s_i s_j\rangle\langle  s_j\rangle}}
+Nq^2\sum_{ij}{\overline{\langle  s_i s_j\rangle^2}}
 - 3q^4N^3-NG_1G_2-2NG_3^2
\\  
\Y_{15} &=&N^3\langle\delta Q_{ab} \delta Q_{bc} \delta Q_{de} \delta Q_{ef}  \rangle_c
\\
\nonumber
&=& \frac{1}{N}\sum_{ijkl}{\overline{\langle  s_i\rangle\langle  s_is_j\rangle\langle  s_j\rangle\langle  s_k\rangle\langle  s_ks_l\rangle\langle  s_l\rangle}}
 -4 q  \sum_{ijk}{\overline{\langle s_i \rangle^2\langle s_j \rangle\langle s_js_k \rangle\langle s_k \rangle}} 
 \\
 \nonumber
 &&
 +4N q^2 \sum_{ij}{\overline{\langle  s_i\rangle^2\langle  s_j\rangle^2}}
 +2N q^2\sum_{ij}{\overline{\langle  s_i\rangle\langle  s_i s_j\rangle\langle  s_j\rangle}}
 - 3q^4N^3-NG_2^2-2NG_3^2
\\
\Y_{16} &=&N^3\langle\delta Q_{ab}^3 \delta Q_{cd}  \rangle_c
\\
\nonumber
&=& \frac{1}{N}\sum_{ijkl}{\overline{\langle  s_is_js_k \rangle^2\langle s_l \rangle^2}}
 -q \sum_{ijk}{\overline{\langle  s_is_js_k\rangle^2}}
 -3q\sum_{ijk}{\overline{\langle s_is_j  \rangle^2\langle s_k \rangle^2}}
 \\
 \nonumber
 &&
 +3Nq^2\sum_{ij}{\overline{\langle  s_i\rangle^2\langle  s_j\rangle^2}}
 +3Nq^2\sum_{ij}{\overline{\langle  s_i s_j\rangle^2}}
 - 3q^4N^3-3NG_1G_3
\end{eqnarray}
\newpage
\begin{eqnarray}
\Y_{17} &=&N^3\langle\delta Q_{ab} \delta Q_{bc}  \delta Q_{ca} \delta Q_{de}\rangle_c
\\
\nonumber
&=& \frac{1}{N}\sum_{ijkl}{\overline{\langle  s_is_j \rangle\langle s_js_k \rangle\langle s_ks_i \rangle\langle s_l \rangle^2}}
 -3q \sum_{ijk}{\overline{\langle s_i \rangle^2\langle s_j \rangle\langle s_js_k \rangle\langle s_k \rangle}} 
  -q \sum_{ijk}{\overline{\langle s_i s_j \rangle\langle s_j s_k \rangle\langle s_k s_i \rangle}}
  \\
  \nonumber
  &&
 +3Nq^2 \sum_{ij}{\overline{\langle  s_i\rangle^2\langle  s_j\rangle^2}}
 +3Nq^2 \sum_{ij}{\overline{\langle  s_i\rangle\langle  s_i s_j\rangle\langle  s_j\rangle}}
 - 3q^4N^3-3NG_2G_3
\\
\Y_{18} &=&N^3\langle\delta Q_{ab}^2 \delta Q _{bc} \delta Q_{de}\rangle_c
\\
\nonumber
&=& \frac{1}{N}\sum_{ijkl}{\overline{\langle  s_is_j\rangle\langle  s_is_js_k\rangle\langle  s_k\rangle\langle  s_l\rangle^2}}
 -2q \sum_{ijk}{\overline{\langle s_i \rangle^2\langle s_j \rangle\langle s_js_k \rangle\langle s_k \rangle}} 
 \\
 \nonumber
 &&
 -q \sum_{ijk}{\overline{\langle s_is_j  \rangle^2\langle s_k \rangle^2}}
 -q  \sum_{ijk}{\overline{\langle s_is_j \rangle\langle s_is_js_k \rangle\langle s_k \rangle}}
 \\
 \nonumber
 &&
 +3Nq^2  \sum_{ij}{\overline{\langle  s_i\rangle^2\langle  s_j\rangle^2}}
 +2Nq^2\sum_{ij}{\overline{\langle  s_i\rangle\langle  s_i s_j\rangle\langle  s_j\rangle}}
 +Nq^2\sum_{ij}{\overline{\langle  s_i s_j\rangle^2}}
 - 3q^4N^3-NG_1G_3 -2NG_2G_3
\\
\Y_{19} &=&N^3\langle\delta Q_{ab} \delta Q_{bc} \delta Q_{cd} \delta Q_{ef}  \rangle_c
\\
\nonumber
&=& \frac{1}{N}\sum_{ijkl}{\overline{\langle s_i  \rangle\langle s_is_j \rangle\langle s_js_k \rangle\langle s_k \rangle\langle s_l \rangle^2}}
 -2q  \sum_{ijk}{\overline{\langle s_i \rangle^2\langle s_j \rangle\langle s_js_k \rangle\langle s_k \rangle}} 
 \\
 &&
 \nonumber
 - q  \sum_{ijk}{\overline{\langle s_i \rangle^2\langle s_j \rangle^2\langle s_k  \rangle^2}}
 - q \sum_{ijk}{\overline{\langle s_i \rangle\langle s_is_j \rangle\langle s_js_k \rangle\langle s_k \rangle}}
 \\
 \nonumber
 &&
 +4 N q^2  \sum_{ij}{\overline{\langle  s_i\rangle^2\langle  s_j\rangle^2}}
 +2 N q^2  \sum_{ij}{\overline{\langle  s_i\rangle\langle  s_i s_j\rangle\langle  s_j\rangle}}
 - 3q^4N^3-2NG_2G_3 -NG3^2
\\
\Y_{20} &=&N^3\langle\delta Q_{ab} \delta Q_{ac} \delta Q_{ad} \delta Q_{ef} \rangle_c
\\
\nonumber
&=& \frac{1}{N}\sum_{ijkl}{\overline{\langle  s_is_js_k\rangle\langle  s_i\rangle\langle  s_j\rangle\langle  s_k\rangle\langle  s_l\rangle^2}}
-3q \sum_{ijk}{\overline{\langle s_i \rangle^2\langle s_j \rangle\langle s_js_k \rangle\langle s_k \rangle}} 
-q \sum_{ijk}{\overline{\langle s_i s_js_k\rangle\langle s_i \rangle\langle s_j \rangle\langle s_k \rangle}} 
\\
\nonumber
&&+3N q^2 \sum_{ij}{\overline{\langle  s_i\rangle\langle  s_i s_j\rangle\langle  s_j\rangle}}
 +3 N q^2 \sum_{ij}{\overline{\langle  s_i\rangle^2\langle  s_j\rangle^2}}
 - 3q^4N^3-3NG_2G_3
\\
\Y_{21} &=&N^3\langle\delta Q_{ab} \delta Q_{cd} \delta Q_{ef}^2  \rangle_c
\\
\nonumber
&=& \frac{1}{N}\sum_{ijkl}{\overline{\langle s_i \rangle^2\langle s_j \rangle^2\langle s_ks_l \rangle^2}}
 -2q \sum_{ijk}{\overline{\langle s_i \rangle^2\langle s_js_k \rangle^2}}
 -2q \sum_{ijk}{\overline{\langle s_i \rangle^2\langle s_j \rangle^2\langle s_k  \rangle^2}}
 \\
\nonumber
&& +5 N q^2  \sum_{ij}{\overline{\langle s_i \rangle^2\langle s_j \rangle^2}}
+ N q^2 \sum_{ij}{\overline{\langle s_is_j  \rangle^2}}
 - 3q^4N^3-NG_1G_3-2NG_3^2
\\
\Y_{22} &=&N^3\langle\delta Q_{ab} \delta Q_{cd} \delta Q_{ef} \delta Q_{fg}  \rangle_c
\\
\nonumber
&=& \frac{1}{N}\sum_{ijkl}{\overline{\langle s_i \rangle^2 \langle s_j \rangle^2\langle s_k \rangle\langle s_ks_l  \rangle\langle s_l \rangle}}
 -2q \sum_{ijk}{\overline{\langle s_i \rangle^2\langle s_j \rangle\langle s_js_k \rangle\langle s_k \rangle}} 
 -2 q \sum_{ijk}{\overline{\langle s_i \rangle^2\langle s_j \rangle^2\langle s_k  \rangle^2}}
 \\
\nonumber
&& +N q^2 \sum_{ij}{\overline{\langle  s_i\rangle\langle  s_i s_j\rangle\langle  s_j\rangle}}
 +5 N q^2 \sum_{ij}{\overline{\langle  s_i\rangle^2\langle  s_j\rangle^2}}
 - 3q^4N^3-NG_2G_3-2NG_3^2
\\
\Y_{23} &=&N^3\langle\delta Q_{ab} \delta Q_{cd} \delta Q_{ef} \delta Q_{gh} \rangle
\\
\nonumber
&=& \frac{1}{N}\sum_{ijkl}{\overline{\langle s_i \rangle^2\langle s_j \rangle^2\langle s_k \rangle^2\langle s_l \rangle^2}}
 -4 q \sum_{ijk}{\overline{\langle s_i \rangle^2\langle s_j \rangle^2\langle s_k  \rangle^2}}
 +6 N q^2 \sum_{ij}{\overline{\langle  s_i\rangle^2\langle  s_j\rangle^2}}
 - 3q^4N^3-3NG_3^2
 \label{Y23}
\end{eqnarray}

Counting the multiplicity of each apart term, the above coefficients in Eq. (\ref{quarcon}) recombine according to the following formulas (they are equal to those of Appendix B in 
\cite{temesvari2007replica} taking into account for the different naming convention).
\begin{eqnarray}
\y_{1} &=& 3 (\Y_1 - 4 \Y_{12} + 2 \Y_{15} + 4 \Y_{19} - 4 \Y_{22} + \Y_{23})  
    \label{y1}
\\   
\y_{2} &=& \frac{1}{2}
  (6 \Y_1 - 12 \Y_{10} - 48 \Y_{11} - 48 \Y_{12} + 3 \Y_{13} - 24 \Y_{14} + 60 \Y_{15} + 
    4 \Y_{16} - 48 \Y_{18}
    \label{y2}
    \\
    \nonumber
    && + 96 \Y_{19} + \Y_2 + 48 \Y_{20} + 24 \Y_{21} - 144 \Y_{22} + 
    36 \Y_{23} - 6 \Y_3 + 12 \Y_5 - 8 \Y_7 + 24 \Y_8 + 24 \Y_9)  
    \\   
\y_{3} &=& 3 (-2 \Y_1 + \Y_{10} + 4 \Y_{11} + 12 \Y_{12} - \Y_{13} + 6 \Y_{14} - 13 \Y_{15} + 
    4 \Y_{18} 
    \label{y3}
    \\
    \nonumber
    &&- 16 \Y_{19} - 4 \Y_{20} - 4 \Y_{21} + 24 \Y_{22} - 6 \Y_{23} + \Y_3 - 2 \Y_8 - 
    4 \Y_9)  \\  
     \y_{4} &=& 
 6 (8 \Y_{11} + 6 \Y_{12} + \Y_{14} - 6 \Y_{15} + 2 \Y_{17} + 2 \Y_{18} - 14 \Y_{19} 
    \label{y4}
 \\
 \nonumber
 &&- 4 \Y_{20} - 
    \Y_{21} + 16 \Y_{22} - 4 \Y_{23} + \Y_4 - \Y_5 - 4 \Y_6 - 2 \Y_9)
      \\   
\y_{5} &=& 6 (-4 \Y_{11} + 2 \Y_{15} - 2 \Y_{18} + 6 \Y_{19} + 4 \Y_{20} + \Y_{21} - 12 \Y_{22} + 
    4 \Y_{23} + \Y_5)  
       \label{y5}
 \\   
\y_{6} &=& 12 (-2 \Y_{11} - 2 \Y_{12} + 2 \Y_{15} - \Y_{17} + 6 \Y_{19} + \Y_{20} - 7 \Y_{22} + 
    2 \Y_{23} + \Y_6)  
    \label{y6}
    \\   
\y_{7} &=& 4 (2 \Y_{10} + 12 \Y_{11} + 6 \Y_{12} + 3 \Y_{14} - 12 \Y_{15} - \Y_{16} + 12 \Y_{18} - 
    24 \Y_{19} 
    \label{y7}
    \\
    \nonumber
    &&- 14 \Y_{20} - 6 \Y_{21} + 42 \Y_{22} - 12 \Y_{23} - 3 \Y_5 + \Y_7 - 3 \Y_8 - 
    3 \Y_9)  \\   
    \y_{8} &=& 
 6 (-\Y_{10} - 2 \Y_{11} - 2 \Y_{12} - \Y_{14} + 5 \Y_{15} - 2 \Y_{18} + 8 \Y_{19} + 4 \Y_{20} + 
    2 \Y_{21} - 18 \Y_{22} + 6 \Y_{23} + \Y_8)  
      \label{y8}
  \\   
\y_{9} &=& 12 (-\Y_{11} - 2 \Y_{12} - \Y_{14} + 3 \Y_{15} - \Y_{18} + 4 \Y_{19} + \Y_{20} + \Y_{21} - 
    7 \Y_{22} + 2 \Y_{23} + \Y_9)  
        \label{y9}
\\   
\y_{10} &=& \Y_{10} - 3 \Y_{15} - 4 \Y_{20} + 12 \Y_{22} - 6 \Y_{23} 
    \label{y10}
\\
\y_{11} &=& 12 (\Y_{11} - \Y_{15} - 2 \Y_{19} - \Y_{20} + 5 \Y_{22} - 2 \Y_{23})  
    \label{y11}
\\   
\y_{12} &=& 12 (\Y_{12} - \Y_{15} - 2 \Y_{19} + 3 \Y_{22} - \Y_{23})  
    \label{y12}
\\   
\y_{13} &=& \frac{3}{4} (\Y_{13} - 4 \Y_{14} + 4 \Y_{15} + 2 \Y_{21} - 4 \Y_{22} + \Y_{23})  
    \label{y13}
\\   
\y_{14} &=& 3 (\Y_{14} - 2 \Y_{15} - \Y_{21} + 3 \Y_{22} - \Y_{23})  
    \label{y14}
\\   
\y_{15} &=& 3 (\Y_{15} - 2 \Y_{22} + \Y_{23})  
    \label{y15}
\\   
\y_{16} &=& \Y_{16} - 6 \Y_{18} + 6 \Y_{19} + 4 \Y_{20} + 3 \Y_{21} - 12 \Y_{22} + 4 \Y_{23}
    \label{y16}
\\
\y_{17} &=& 2 (\Y_{17} - 3 \Y_{19} + 3 \Y_{22} - \Y_{23})  
    \label{y17}
\\   
\y_{18} &=& 6 (\Y_{18} - 2 \Y_{19} - \Y_{20} - \Y_{21} + 5 \Y_{22} - 2 \Y_{23}) 
     \label{y18}
\\   
\y_{19} &=& 6 (\Y_{19} - 2 \Y_{22} + \Y_{23}) 
    \label{y19}
 \\ 
  \y_{20} &=& 2 (\Y_{20} - 3 \Y_{22} + 2 \Y_{23}) 
       \label{y20}
\\   
\y_{21} &=&\frac{ 3}{4} (\Y_{21} - 2 \Y_{22} + \Y_{23}) 
    \label{y21}
 \\   
\y_{22} &=& \frac{3}{2} (\Y_{22} - \Y_{23})
    \label{y22}
 \\
\y_{23} &=& \frac{1}{16}\Y_{23}
    \label{y23}
\end{eqnarray}

\bibliography{lambdauno}

\begin{thebibliography}{58}
\expandafter\ifx\csname natexlab\endcsname\relax\def\natexlab#1{#1}\fi
\expandafter\ifx\csname bibnamefont\endcsname\relax
  \def\bibnamefont#1{#1}\fi
\expandafter\ifx\csname bibfnamefont\endcsname\relax
  \def\bibfnamefont#1{#1}\fi
\expandafter\ifx\csname citenamefont\endcsname\relax
  \def\citenamefont#1{#1}\fi
\expandafter\ifx\csname url\endcsname\relax
  \def\url#1{\texttt{#1}}\fi
\expandafter\ifx\csname urlprefix\endcsname\relax\def\urlprefix{URL }\fi
\providecommand{\bibinfo}[2]{#2}
\providecommand{\eprint}[2][]{\url{#2}}

\bibitem[{\citenamefont{{G\"otze}}(1984)}]{Goetze84}
\bibinfo{author}{\bibfnamefont{W.}~\bibnamefont{{G\"otze}}},
  \bibinfo{journal}{Z. Phys. B} \textbf{\bibinfo{volume}{{\bf 56}}},
  \bibinfo{pages}{139} (\bibinfo{year}{1984}).

\bibitem[{\citenamefont{{G\"otze}}(1985)}]{Goetze85}
\bibinfo{author}{\bibfnamefont{W.}~\bibnamefont{{G\"otze}}},
  \bibinfo{journal}{Z. Phys. B} \textbf{\bibinfo{volume}{{\bf 60}}},
  \bibinfo{pages}{195} (\bibinfo{year}{1985}).

\bibitem[{\citenamefont{{G\"otze}}(1991)}]{Goetze89}
\bibinfo{author}{\bibfnamefont{W.}~\bibnamefont{{G\"otze}}}, in
  \emph{\bibinfo{booktitle}{Les Houches Session 1989}}, edited by
  \bibinfo{editor}{\bibfnamefont{J.}~\bibnamefont{Hansen}},
  \bibinfo{editor}{\bibfnamefont{D.}~\bibnamefont{Levesque}}, \bibnamefont{and}
  \bibinfo{editor}{\bibfnamefont{J.}~\bibnamefont{Zinn-Justin}}
  (\bibinfo{publisher}{North Holland (Amsterdam)}, \bibinfo{year}{1991}).

\bibitem[{\citenamefont{{G\"otze}}(2009)}]{Goetze09}
\bibinfo{author}{\bibfnamefont{W.}~\bibnamefont{{G\"otze}}},
  \emph{\bibinfo{title}{Complex Dynamics of Glass-Forming Liquids: A
  Mode-Coupling Theory}} (\bibinfo{publisher}{OUP (Oxford, UK)},
  \bibinfo{year}{2009}).

\bibitem[{\citenamefont{Gotze and Sjogren}(1989)}]{Goetze89c}
\bibinfo{author}{\bibfnamefont{W.}~\bibnamefont{Gotze}} \bibnamefont{and}
  \bibinfo{author}{\bibfnamefont{L.}~\bibnamefont{Sjogren}},
  \bibinfo{journal}{Journal of Physics: Condensed Matter}
  \textbf{\bibinfo{volume}{1}}, \bibinfo{pages}{4203} (\bibinfo{year}{1989}),
  \urlprefix\url{https://dx.doi.org/10.1088/0953-8984/1/26/015}.

\bibitem[{\citenamefont{Crisanti et~al.}(1993)\citenamefont{Crisanti, Horner,
  and Sommers}}]{crisanti1993spherical}
\bibinfo{author}{\bibfnamefont{A.}~\bibnamefont{Crisanti}},
  \bibinfo{author}{\bibfnamefont{H.}~\bibnamefont{Horner}}, \bibnamefont{and}
  \bibinfo{author}{\bibfnamefont{H.~J.} \bibnamefont{Sommers}},
  \bibinfo{journal}{Zeitschrift f{\"u}r Physik B Condensed Matter}
  \textbf{\bibinfo{volume}{92}}, \bibinfo{pages}{257} (\bibinfo{year}{1993}).

\bibitem[{\citenamefont{Crisanti and Leuzzi}(2006)}]{Crisanti06}
\bibinfo{author}{\bibfnamefont{A.}~\bibnamefont{Crisanti}} \bibnamefont{and}
  \bibinfo{author}{\bibfnamefont{L.}~\bibnamefont{Leuzzi}},
  \bibinfo{journal}{Phys. Rev. B} \textbf{\bibinfo{volume}{{\bf 73}}},
  \bibinfo{pages}{014412} (\bibinfo{year}{2006}).

\bibitem[{\citenamefont{Crisanti and Leuzzi}(2015)}]{Crisanti15}
\bibinfo{author}{\bibfnamefont{A.}~\bibnamefont{Crisanti}} \bibnamefont{and}
  \bibinfo{author}{\bibfnamefont{L.}~\bibnamefont{Leuzzi}},
  \bibinfo{journal}{Journal of Non-Crystalline Solids}
  \textbf{\bibinfo{volume}{407}}, \bibinfo{pages}{110} (\bibinfo{year}{2015}),
  ISSN \bibinfo{issn}{0022-3093}, \bibinfo{note}{7th IDMRCS: Relaxation in
  Complex Systems},
  \urlprefix\url{https://www.sciencedirect.com/science/article/pii/S0022309314003640}.

\bibitem[{\citenamefont{Franz et~al.}(2013)\citenamefont{Franz, Parisi,
  Ricci-Tersenghi, Rizzo, and Urbani}}]{franz2013note}
\bibinfo{author}{\bibfnamefont{S.}~\bibnamefont{Franz}},
  \bibinfo{author}{\bibfnamefont{G.}~\bibnamefont{Parisi}},
  \bibinfo{author}{\bibfnamefont{F.}~\bibnamefont{Ricci-Tersenghi}},
  \bibinfo{author}{\bibfnamefont{T.}~\bibnamefont{Rizzo}}, \bibnamefont{and}
  \bibinfo{author}{\bibfnamefont{P.}~\bibnamefont{Urbani}},
  \bibinfo{journal}{The Journal of Chemical Physics}
  \textbf{\bibinfo{volume}{138}} (\bibinfo{year}{2013}).

\bibitem[{\citenamefont{Cammarota and
  Biroli}(2012{\natexlab{a}})}]{cammarota2012aging}
\bibinfo{author}{\bibfnamefont{C.}~\bibnamefont{Cammarota}} \bibnamefont{and}
  \bibinfo{author}{\bibfnamefont{G.}~\bibnamefont{Biroli}},
  \bibinfo{journal}{Europhysics Letters} \textbf{\bibinfo{volume}{98}},
  \bibinfo{pages}{16011} (\bibinfo{year}{2012}{\natexlab{a}}).

\bibitem[{\citenamefont{Krakoviack}(2005)}]{Krakoviack05}
\bibinfo{author}{\bibfnamefont{V.}~\bibnamefont{Krakoviack}},
  \bibinfo{journal}{Physical Review Letters} \textbf{\bibinfo{volume}{94}},
  \bibinfo{pages}{65703} (\bibinfo{year}{2005}).

\bibitem[{\citenamefont{Krakoviack}(2007)}]{Krakoviack07}
\bibinfo{author}{\bibfnamefont{V.}~\bibnamefont{Krakoviack}},
  \bibinfo{journal}{Physical Review E} \textbf{\bibinfo{volume}{75}},
  \bibinfo{pages}{031503} (\bibinfo{year}{2007}).

\bibitem[{\citenamefont{Madden and Glandt}(1988)}]{Madden88}
\bibinfo{author}{\bibfnamefont{W.}~\bibnamefont{Madden}} \bibnamefont{and}
  \bibinfo{author}{\bibfnamefont{E.}~\bibnamefont{Glandt}},
  \bibinfo{journal}{Journal of Statistical Physics}
  \textbf{\bibinfo{volume}{51}}, \bibinfo{pages}{537} (\bibinfo{year}{1988}).

\bibitem[{\citenamefont{Given and Stell}(1992)}]{Given92}
\bibinfo{author}{\bibfnamefont{J.}~\bibnamefont{Given}} \bibnamefont{and}
  \bibinfo{author}{\bibfnamefont{G.}~\bibnamefont{Stell}},
  \bibinfo{journal}{The Journal of Chemical Physics}
  \textbf{\bibinfo{volume}{97}}, \bibinfo{pages}{4573} (\bibinfo{year}{1992}).

\bibitem[{\citenamefont{Cammarota and
  Biroli}(2012{\natexlab{b}})}]{cammarota2012ideal}
\bibinfo{author}{\bibfnamefont{C.}~\bibnamefont{Cammarota}} \bibnamefont{and}
  \bibinfo{author}{\bibfnamefont{G.}~\bibnamefont{Biroli}},
  \bibinfo{journal}{Proceedings of the National Academy of Sciences}
  \textbf{\bibinfo{volume}{109}}, \bibinfo{pages}{8850}
  (\bibinfo{year}{2012}{\natexlab{b}}).

\bibitem[{\citenamefont{Sjogren}(1991)}]{Sjogren91}
\bibinfo{author}{\bibfnamefont{L.}~\bibnamefont{Sjogren}},
  \bibinfo{journal}{Journal of Physics: Condensed Matter}
  \textbf{\bibinfo{volume}{3}}, \bibinfo{pages}{5023} (\bibinfo{year}{1991}),
  \urlprefix\url{https://dx.doi.org/10.1088/0953-8984/3/26/021}.

\bibitem[{\citenamefont{Zaccarelli et~al.}(2002)\citenamefont{Zaccarelli,
  Foffi, Dawson, Buldyrev, Sciortino, and Tartaglia}}]{Zaccarelli02}
\bibinfo{author}{\bibfnamefont{E.}~\bibnamefont{Zaccarelli}},
  \bibinfo{author}{\bibfnamefont{G.}~\bibnamefont{Foffi}},
  \bibinfo{author}{\bibfnamefont{K.~A.} \bibnamefont{Dawson}},
  \bibinfo{author}{\bibfnamefont{S.~V.} \bibnamefont{Buldyrev}},
  \bibinfo{author}{\bibfnamefont{F.}~\bibnamefont{Sciortino}},
  \bibnamefont{and}
  \bibinfo{author}{\bibfnamefont{P.}~\bibnamefont{Tartaglia}},
  \bibinfo{journal}{Phys. Rev. E} \textbf{\bibinfo{volume}{66}},
  \bibinfo{pages}{041402} (\bibinfo{year}{2002}),
  \urlprefix\url{https://link.aps.org/doi/10.1103/PhysRevE.66.041402}.

\bibitem[{\citenamefont{Sciortino
  et~al.}(2003{\natexlab{a}})\citenamefont{Sciortino, La~Nave, and
  Tartaglia}}]{Sciortino03}
\bibinfo{author}{\bibfnamefont{F.}~\bibnamefont{Sciortino}},
  \bibinfo{author}{\bibfnamefont{E.}~\bibnamefont{La~Nave}}, \bibnamefont{and}
  \bibinfo{author}{\bibfnamefont{P.}~\bibnamefont{Tartaglia}},
  \bibinfo{journal}{Phys. Rev. Lett.} \textbf{\bibinfo{volume}{{\bf 91}}},
  \bibinfo{pages}{155701} (\bibinfo{year}{2003}{\natexlab{a}}).

\bibitem[{\citenamefont{Sciortino
  et~al.}(2003{\natexlab{b}})\citenamefont{Sciortino, Tartaglia, and
  Zaccarelli}}]{Sciortino03b}
\bibinfo{author}{\bibfnamefont{F.}~\bibnamefont{Sciortino}},
  \bibinfo{author}{\bibfnamefont{P.}~\bibnamefont{Tartaglia}},
  \bibnamefont{and}
  \bibinfo{author}{\bibfnamefont{E.}~\bibnamefont{Zaccarelli}},
  \bibinfo{journal}{Phys. Rev. Lett.} \textbf{\bibinfo{volume}{91}},
  \bibinfo{pages}{268301} (\bibinfo{year}{2003}{\natexlab{b}}),
  \urlprefix\url{https://link.aps.org/doi/10.1103/PhysRevLett.91.268301}.

\bibitem[{\citenamefont{Caltagirone
  et~al.}(2012{\natexlab{a}})\citenamefont{Caltagirone, Ferrari, Leuzzi,
  Parisi, Ricci-Tersenghi, and Rizzo}}]{Caltagirone12a}
\bibinfo{author}{\bibfnamefont{F.}~\bibnamefont{Caltagirone}},
  \bibinfo{author}{\bibfnamefont{U.}~\bibnamefont{Ferrari}},
  \bibinfo{author}{\bibfnamefont{L.}~\bibnamefont{Leuzzi}},
  \bibinfo{author}{\bibfnamefont{G.}~\bibnamefont{Parisi}},
  \bibinfo{author}{\bibfnamefont{F.}~\bibnamefont{Ricci-Tersenghi}},
  \bibnamefont{and} \bibinfo{author}{\bibfnamefont{T.}~\bibnamefont{Rizzo}},
  \bibinfo{journal}{Phys. Rev. Lett.} \textbf{\bibinfo{volume}{108}},
  \bibinfo{pages}{085702} (\bibinfo{year}{2012}{\natexlab{a}}),
  \urlprefix\url{https://link.aps.org/doi/10.1103/PhysRevLett.108.085702}.

\bibitem[{\citenamefont{Caltagirone
  et~al.}(2012{\natexlab{b}})\citenamefont{Caltagirone, Ferrari, Leuzzi,
  Parisi, and Rizzo}}]{Caltagirone12b}
\bibinfo{author}{\bibfnamefont{F.}~\bibnamefont{Caltagirone}},
  \bibinfo{author}{\bibfnamefont{U.}~\bibnamefont{Ferrari}},
  \bibinfo{author}{\bibfnamefont{L.}~\bibnamefont{Leuzzi}},
  \bibinfo{author}{\bibfnamefont{G.}~\bibnamefont{Parisi}}, \bibnamefont{and}
  \bibinfo{author}{\bibfnamefont{T.}~\bibnamefont{Rizzo}},
  \bibinfo{journal}{Phys. Rev. B} \textbf{\bibinfo{volume}{86}},
  \bibinfo{pages}{064204} (\bibinfo{year}{2012}{\natexlab{b}}),
  \urlprefix\url{https://link.aps.org/doi/10.1103/PhysRevB.86.064204}.

\bibitem[{\citenamefont{Caltagirone
  et~al.}(2012{\natexlab{c}})\citenamefont{Caltagirone, Parisi, and
  Rizzo}}]{Caltagirone12c}
\bibinfo{author}{\bibfnamefont{F.}~\bibnamefont{Caltagirone}},
  \bibinfo{author}{\bibfnamefont{G.}~\bibnamefont{Parisi}}, \bibnamefont{and}
  \bibinfo{author}{\bibfnamefont{T.}~\bibnamefont{Rizzo}},
  \bibinfo{journal}{Phys. Rev. E} \textbf{\bibinfo{volume}{85}},
  \bibinfo{pages}{051504} (\bibinfo{year}{2012}{\natexlab{c}}),
  \urlprefix\url{https://link.aps.org/doi/10.1103/PhysRevE.85.051504}.

\bibitem[{\citenamefont{Ferrari et~al.}(2012)\citenamefont{Ferrari, Leuzzi,
  Parisi, and Rizzo}}]{Ferrari12}
\bibinfo{author}{\bibfnamefont{U.}~\bibnamefont{Ferrari}},
  \bibinfo{author}{\bibfnamefont{L.}~\bibnamefont{Leuzzi}},
  \bibinfo{author}{\bibfnamefont{G.}~\bibnamefont{Parisi}}, \bibnamefont{and}
  \bibinfo{author}{\bibfnamefont{T.}~\bibnamefont{Rizzo}},
  \bibinfo{journal}{Phys. Rev. B} \textbf{\bibinfo{volume}{86}},
  \bibinfo{pages}{014204} (\bibinfo{year}{2012}),
  \urlprefix\url{https://link.aps.org/doi/10.1103/PhysRevB.86.014204}.

\bibitem[{\citenamefont{Parisi and Rizzo}(2013)}]{Parisi13a}
\bibinfo{author}{\bibfnamefont{G.}~\bibnamefont{Parisi}} \bibnamefont{and}
  \bibinfo{author}{\bibfnamefont{T.}~\bibnamefont{Rizzo}},
  \bibinfo{journal}{Phys. Rev. E} \textbf{\bibinfo{volume}{87}},
  \bibinfo{pages}{012101} (\bibinfo{year}{2013}),
  \urlprefix\url{https://link.aps.org/doi/10.1103/PhysRevE.87.012101}.

\bibitem[{\citenamefont{M\'{e}zard and Parisi}(1996)}]{Mezard96}
\bibinfo{author}{\bibfnamefont{M.}~\bibnamefont{M\'{e}zard}} \bibnamefont{and}
  \bibinfo{author}{\bibfnamefont{G.}~\bibnamefont{Parisi}},
  \bibinfo{journal}{J. Phys. A: Math. and Gen.} \textbf{\bibinfo{volume}{{\bf
  29}}}, \bibinfo{pages}{6515} (\bibinfo{year}{1996}).

\bibitem[{\citenamefont{M\'ezard and Parisi}(1999{\natexlab{a}})}]{Mezard99a}
\bibinfo{author}{\bibfnamefont{M.}~\bibnamefont{M\'ezard}} \bibnamefont{and}
  \bibinfo{author}{\bibfnamefont{G.}~\bibnamefont{Parisi}},
  \bibinfo{journal}{Phys. Rev. Lett.} \textbf{\bibinfo{volume}{{\bf 82}}},
  \bibinfo{pages}{747} (\bibinfo{year}{1999}{\natexlab{a}}).

\bibitem[{\citenamefont{M\'ezard}(1999)}]{Mezard99b}
\bibinfo{author}{\bibfnamefont{M.}~\bibnamefont{M\'ezard}},
  \bibinfo{journal}{Physica A} \textbf{\bibinfo{volume}{{\bf 265}}},
  \bibinfo{pages}{352} (\bibinfo{year}{1999}).

\bibitem[{\citenamefont{M\'ezard and Parisi}(1999{\natexlab{b}})}]{Mezard99c}
\bibinfo{author}{\bibfnamefont{M.}~\bibnamefont{M\'ezard}} \bibnamefont{and}
  \bibinfo{author}{\bibfnamefont{G.}~\bibnamefont{Parisi}},
  \bibinfo{journal}{J. Chem. Phys.} \textbf{\bibinfo{volume}{{\bf 111}}},
  \bibinfo{pages}{1076} (\bibinfo{year}{1999}{\natexlab{b}}).

\bibitem[{\citenamefont{M\'ezard and Parisi}(2000)}]{Mezard00}
\bibinfo{author}{\bibfnamefont{M.}~\bibnamefont{M\'ezard}} \bibnamefont{and}
  \bibinfo{author}{\bibfnamefont{G.}~\bibnamefont{Parisi}},
  \bibinfo{journal}{J. Phys.: Cond. Matt.} \textbf{\bibinfo{volume}{{\bf 12}}},
  \bibinfo{pages}{6655} (\bibinfo{year}{2000}).

\bibitem[{\citenamefont{Parisi et~al.}(2020)\citenamefont{Parisi, Urbani, and
  Zamponi}}]{parisi2020theory}
\bibinfo{author}{\bibfnamefont{G.}~\bibnamefont{Parisi}},
  \bibinfo{author}{\bibfnamefont{P.}~\bibnamefont{Urbani}}, \bibnamefont{and}
  \bibinfo{author}{\bibfnamefont{F.}~\bibnamefont{Zamponi}},
  \emph{\bibinfo{title}{Theory of simple glasses: exact solutions in infinite
  dimensions}} (\bibinfo{publisher}{Cambridge University Press},
  \bibinfo{year}{2020}).

\bibitem[{\citenamefont{Rizzo}(2016)}]{rizzo2016dynamical}
\bibinfo{author}{\bibfnamefont{T.}~\bibnamefont{Rizzo}},
  \bibinfo{journal}{Physical Review B} \textbf{\bibinfo{volume}{94}},
  \bibinfo{pages}{014202} (\bibinfo{year}{2016}).

\bibitem[{\citenamefont{M\'ezard et~al.}(1987)\citenamefont{M\'ezard, Parisi,
  and Virasoro}}]{Mezard87}
\bibinfo{author}{\bibfnamefont{M.}~\bibnamefont{M\'ezard}},
  \bibinfo{author}{\bibfnamefont{G.}~\bibnamefont{Parisi}}, \bibnamefont{and}
  \bibinfo{author}{\bibfnamefont{M.}~\bibnamefont{Virasoro}},
  \emph{\bibinfo{title}{Spin Glass Theory and Beyond}}
  (\bibinfo{publisher}{World Scientific (Singapore)}, \bibinfo{year}{1987}).

\bibitem[{\citenamefont{Franz et~al.}(2011)\citenamefont{Franz, Parisi,
  Ricci-Tersenghi, and Rizzo}}]{franz2011field}
\bibinfo{author}{\bibfnamefont{S.}~\bibnamefont{Franz}},
  \bibinfo{author}{\bibfnamefont{G.}~\bibnamefont{Parisi}},
  \bibinfo{author}{\bibfnamefont{F.}~\bibnamefont{Ricci-Tersenghi}},
  \bibnamefont{and} \bibinfo{author}{\bibfnamefont{T.}~\bibnamefont{Rizzo}},
  \bibinfo{journal}{The European Physical Journal E}
  \textbf{\bibinfo{volume}{34}}, \bibinfo{pages}{1} (\bibinfo{year}{2011}).

\bibitem[{\citenamefont{Crisanti and Sommers}(1992)}]{crisanti1992spherical}
\bibinfo{author}{\bibfnamefont{A.}~\bibnamefont{Crisanti}} \bibnamefont{and}
  \bibinfo{author}{\bibfnamefont{H.-J.} \bibnamefont{Sommers}},
  \bibinfo{journal}{Zeitschrift f{\"u}r Physik B Condensed Matter}
  \textbf{\bibinfo{volume}{87}}, \bibinfo{pages}{341} (\bibinfo{year}{1992}).

\bibitem[{\citenamefont{Caltagirone et~al.}(2013)\citenamefont{Caltagirone,
  Parisi, and Rizzo}}]{caltagirone2013critical}
\bibinfo{author}{\bibfnamefont{F.}~\bibnamefont{Caltagirone}},
  \bibinfo{author}{\bibfnamefont{G.}~\bibnamefont{Parisi}}, \bibnamefont{and}
  \bibinfo{author}{\bibfnamefont{T.}~\bibnamefont{Rizzo}},
  \bibinfo{journal}{Physical Review E} \textbf{\bibinfo{volume}{87}},
  \bibinfo{pages}{032134} (\bibinfo{year}{2013}).

\bibitem[{\citenamefont{Gross et~al.}(1985)\citenamefont{Gross, Kanter, and
  Sompolinsky}}]{gross1985mean}
\bibinfo{author}{\bibfnamefont{D.~J.} \bibnamefont{Gross}},
  \bibinfo{author}{\bibfnamefont{I.}~\bibnamefont{Kanter}}, \bibnamefont{and}
  \bibinfo{author}{\bibfnamefont{H.}~\bibnamefont{Sompolinsky}},
  \bibinfo{journal}{Physical review letters} \textbf{\bibinfo{volume}{55}},
  \bibinfo{pages}{304} (\bibinfo{year}{1985}).

\bibitem[{\citenamefont{Goldschmidt}(1988)}]{goldschmidt1988potts}
\bibinfo{author}{\bibfnamefont{Y.}~\bibnamefont{Goldschmidt}},
  \bibinfo{journal}{Europhysics Letters} \textbf{\bibinfo{volume}{6}},
  \bibinfo{pages}{7} (\bibinfo{year}{1988}).

\bibitem[{\citenamefont{Kurchan}(1992)}]{kurchan1992supersymmetry}
\bibinfo{author}{\bibfnamefont{J.}~\bibnamefont{Kurchan}},
  \bibinfo{journal}{Journal de Physique I} \textbf{\bibinfo{volume}{2}},
  \bibinfo{pages}{1333} (\bibinfo{year}{1992}).

\bibitem[{\citenamefont{Crisanti}(2008)}]{Crisanti08}
\bibinfo{author}{\bibfnamefont{A.}~\bibnamefont{Crisanti}},
  \bibinfo{journal}{Nuclear Physics B} \textbf{\bibinfo{volume}{796}},
  \bibinfo{pages}{425} (\bibinfo{year}{2008}), ISSN \bibinfo{issn}{0550-3213},
  \urlprefix\url{https://www.sciencedirect.com/science/article/pii/S0550321307009248}.

\bibitem[{\citenamefont{Temesv{\'a}ri et~al.}(2002)\citenamefont{Temesv{\'a}ri,
  De~Dominicis, and Pimentel}}]{temesvari2002generic}
\bibinfo{author}{\bibfnamefont{T.}~\bibnamefont{Temesv{\'a}ri}},
  \bibinfo{author}{\bibfnamefont{C.}~\bibnamefont{De~Dominicis}},
  \bibnamefont{and} \bibinfo{author}{\bibfnamefont{I.}~\bibnamefont{Pimentel}},
  \bibinfo{journal}{The European Physical Journal B-Condensed Matter and
  Complex Systems} \textbf{\bibinfo{volume}{25}}, \bibinfo{pages}{361}
  (\bibinfo{year}{2002}).

\bibitem[{\citenamefont{Temesv{\'a}ri}(2007)}]{temesvari2007replica}
\bibinfo{author}{\bibfnamefont{T.}~\bibnamefont{Temesv{\'a}ri}},
  \bibinfo{journal}{Nuclear Physics B} \textbf{\bibinfo{volume}{772}},
  \bibinfo{pages}{340} (\bibinfo{year}{2007}).

\bibitem[{\citenamefont{Marinari and Parisi}(1992)}]{marinari1992simulated}
\bibinfo{author}{\bibfnamefont{E.}~\bibnamefont{Marinari}} \bibnamefont{and}
  \bibinfo{author}{\bibfnamefont{G.}~\bibnamefont{Parisi}},
  \bibinfo{journal}{Europhysics letters} \textbf{\bibinfo{volume}{19}},
  \bibinfo{pages}{451} (\bibinfo{year}{1992}).

\bibitem[{\citenamefont{Hukushima and Nemoto}(1996)}]{hukushima1996exchange}
\bibinfo{author}{\bibfnamefont{K.}~\bibnamefont{Hukushima}} \bibnamefont{and}
  \bibinfo{author}{\bibfnamefont{K.}~\bibnamefont{Nemoto}},
  \bibinfo{journal}{Journal of the Physical Society of Japan}
  \textbf{\bibinfo{volume}{65}}, \bibinfo{pages}{1604} (\bibinfo{year}{1996}).

\bibitem[{\citenamefont{Ninarello et~al.}(2017)\citenamefont{Ninarello,
  Berthier, and Coslovich}}]{ninarello2017models}
\bibinfo{author}{\bibfnamefont{A.}~\bibnamefont{Ninarello}},
  \bibinfo{author}{\bibfnamefont{L.}~\bibnamefont{Berthier}}, \bibnamefont{and}
  \bibinfo{author}{\bibfnamefont{D.}~\bibnamefont{Coslovich}},
  \bibinfo{journal}{Physical Review X} \textbf{\bibinfo{volume}{7}},
  \bibinfo{pages}{021039} (\bibinfo{year}{2017}).

\bibitem[{\citenamefont{Albert et~al.}(2019)\citenamefont{Albert, Michl,
  Lunkenheimer, Loidl, D{\'e}jardin, and Ladieu}}]{albert2019third}
\bibinfo{author}{\bibfnamefont{S.}~\bibnamefont{Albert}},
  \bibinfo{author}{\bibfnamefont{M.}~\bibnamefont{Michl}},
  \bibinfo{author}{\bibfnamefont{P.}~\bibnamefont{Lunkenheimer}},
  \bibinfo{author}{\bibfnamefont{A.}~\bibnamefont{Loidl}},
  \bibinfo{author}{\bibfnamefont{P.}~\bibnamefont{D{\'e}jardin}},
  \bibnamefont{and} \bibinfo{author}{\bibfnamefont{F.}~\bibnamefont{Ladieu}},
  \bibinfo{journal}{Journal of Statistical Mechanics: Theory and Experiment}
  \textbf{\bibinfo{volume}{2019}}, \bibinfo{pages}{124003}
  (\bibinfo{year}{2019}).

\bibitem[{\citenamefont{Ikeda et~al.}(2017)\citenamefont{Ikeda, Miyazaki, and
  Biroli}}]{ikeda2017fredrickson}
\bibinfo{author}{\bibfnamefont{H.}~\bibnamefont{Ikeda}},
  \bibinfo{author}{\bibfnamefont{K.}~\bibnamefont{Miyazaki}}, \bibnamefont{and}
  \bibinfo{author}{\bibfnamefont{G.}~\bibnamefont{Biroli}},
  \bibinfo{journal}{Europhysics Letters} \textbf{\bibinfo{volume}{116}},
  \bibinfo{pages}{56004} (\bibinfo{year}{2017}).

\bibitem[{\citenamefont{Sellitto et~al.}(2010)\citenamefont{Sellitto,
  De~Martino, Caccioli, and Arenzon}}]{sellitto2010dynamic}
\bibinfo{author}{\bibfnamefont{M.}~\bibnamefont{Sellitto}},
  \bibinfo{author}{\bibfnamefont{D.}~\bibnamefont{De~Martino}},
  \bibinfo{author}{\bibfnamefont{F.}~\bibnamefont{Caccioli}}, \bibnamefont{and}
  \bibinfo{author}{\bibfnamefont{J.~J.} \bibnamefont{Arenzon}},
  \bibinfo{journal}{Physical review letters} \textbf{\bibinfo{volume}{105}},
  \bibinfo{pages}{265704} (\bibinfo{year}{2010}).

\bibitem[{\citenamefont{Arenzon and Sellitto}(2012)}]{arenzon2012microscopic}
\bibinfo{author}{\bibfnamefont{J.~J.} \bibnamefont{Arenzon}} \bibnamefont{and}
  \bibinfo{author}{\bibfnamefont{M.}~\bibnamefont{Sellitto}},
  \bibinfo{journal}{The Journal of chemical physics}
  \textbf{\bibinfo{volume}{137}}, \bibinfo{pages}{084501}
  (\bibinfo{year}{2012}).

\bibitem[{\citenamefont{Perrupato and Rizzo}(2022)}]{perrupato2022exact}
\bibinfo{author}{\bibfnamefont{G.}~\bibnamefont{Perrupato}} \bibnamefont{and}
  \bibinfo{author}{\bibfnamefont{T.}~\bibnamefont{Rizzo}},
  \bibinfo{journal}{arXiv preprint arXiv:2212.05132}  (\bibinfo{year}{2022}).

\bibitem[{\citenamefont{Perrupato and
  Rizzo}(2023)}]{perrupato2023thermodynamics}
\bibinfo{author}{\bibfnamefont{G.}~\bibnamefont{Perrupato}} \bibnamefont{and}
  \bibinfo{author}{\bibfnamefont{T.}~\bibnamefont{Rizzo}},
  \bibinfo{journal}{arXiv preprint arXiv:2312.01430}  (\bibinfo{year}{2023}).

\bibitem[{\citenamefont{Rizzo}(2014)}]{rizzo2014long}
\bibinfo{author}{\bibfnamefont{T.}~\bibnamefont{Rizzo}},
  \bibinfo{journal}{Europhysics Letters} \textbf{\bibinfo{volume}{106}},
  \bibinfo{pages}{56003} (\bibinfo{year}{2014}).

\bibitem[{\citenamefont{Rizzo and Voigtmann}(2015)}]{rizzo2015qualitative}
\bibinfo{author}{\bibfnamefont{T.}~\bibnamefont{Rizzo}} \bibnamefont{and}
  \bibinfo{author}{\bibfnamefont{T.}~\bibnamefont{Voigtmann}},
  \bibinfo{journal}{Europhysics Letters} \textbf{\bibinfo{volume}{111}},
  \bibinfo{pages}{56008} (\bibinfo{year}{2015}).

\bibitem[{\citenamefont{Rizzo and Voigtmann}(2020)}]{rizzo2020solvable}
\bibinfo{author}{\bibfnamefont{T.}~\bibnamefont{Rizzo}} \bibnamefont{and}
  \bibinfo{author}{\bibfnamefont{T.}~\bibnamefont{Voigtmann}},
  \bibinfo{journal}{Physical Review Letters} \textbf{\bibinfo{volume}{124}},
  \bibinfo{pages}{195501} (\bibinfo{year}{2020}).

\bibitem[{\citenamefont{Elizondo-Aguilera
  et~al.}(2022)\citenamefont{Elizondo-Aguilera, Rizzo, and
  Voigtmann}}]{elizondo2022subaging}
\bibinfo{author}{\bibfnamefont{L.~F.} \bibnamefont{Elizondo-Aguilera}},
  \bibinfo{author}{\bibfnamefont{T.}~\bibnamefont{Rizzo}}, \bibnamefont{and}
  \bibinfo{author}{\bibfnamefont{T.}~\bibnamefont{Voigtmann}},
  \bibinfo{journal}{Physical Review Letters} \textbf{\bibinfo{volume}{129}},
  \bibinfo{pages}{238003} (\bibinfo{year}{2022}).

\bibitem[{\citenamefont{Bray and Roberts}(1980)}]{bray1980renormalisation}
\bibinfo{author}{\bibfnamefont{A.}~\bibnamefont{Bray}} \bibnamefont{and}
  \bibinfo{author}{\bibfnamefont{S.}~\bibnamefont{Roberts}},
  \bibinfo{journal}{Journal of Physics C: Solid State Physics}
  \textbf{\bibinfo{volume}{13}}, \bibinfo{pages}{5405} (\bibinfo{year}{1980}).

\bibitem[{\citenamefont{Lubensky et~al.}(2023)\citenamefont{Lubensky,
  Temesv{\'a}ri, Kondor, and Angelini}}]{lubensky2023renormalization}
\bibinfo{author}{\bibfnamefont{T.}~\bibnamefont{Lubensky}},
  \bibinfo{author}{\bibfnamefont{T.}~\bibnamefont{Temesv{\'a}ri}},
  \bibinfo{author}{\bibfnamefont{I.}~\bibnamefont{Kondor}}, \bibnamefont{and}
  \bibinfo{author}{\bibfnamefont{M.~C.} \bibnamefont{Angelini}}, in
  \emph{\bibinfo{booktitle}{Spin Glass Theory and Far Beyond: Replica Symmetry
  Breaking After 40 Years}} (\bibinfo{publisher}{World Scientific},
  \bibinfo{year}{2023}), pp. \bibinfo{pages}{45--67}.

\bibitem[{\citenamefont{Angelini et~al.}(2022)\citenamefont{Angelini,
  Lucibello, Parisi, Perrupato, Ricci-Tersenghi, and
  Rizzo}}]{angelini2022unexpected}
\bibinfo{author}{\bibfnamefont{M.~C.} \bibnamefont{Angelini}},
  \bibinfo{author}{\bibfnamefont{C.}~\bibnamefont{Lucibello}},
  \bibinfo{author}{\bibfnamefont{G.}~\bibnamefont{Parisi}},
  \bibinfo{author}{\bibfnamefont{G.}~\bibnamefont{Perrupato}},
  \bibinfo{author}{\bibfnamefont{F.}~\bibnamefont{Ricci-Tersenghi}},
  \bibnamefont{and} \bibinfo{author}{\bibfnamefont{T.}~\bibnamefont{Rizzo}},
  \bibinfo{journal}{Physical Review Letters} \textbf{\bibinfo{volume}{128}},
  \bibinfo{pages}{075702} (\bibinfo{year}{2022}).

\bibitem[{\citenamefont{Cugliandolo and
  Kurchan}(1993)}]{cugliandolo1993analytical}
\bibinfo{author}{\bibfnamefont{L.~F.} \bibnamefont{Cugliandolo}}
  \bibnamefont{and} \bibinfo{author}{\bibfnamefont{J.}~\bibnamefont{Kurchan}},
  \bibinfo{journal}{Physical Review Letters} \textbf{\bibinfo{volume}{71}},
  \bibinfo{pages}{173} (\bibinfo{year}{1993}).

\end{thebibliography}

\end{document}